\documentclass[a4paper,11pt,amssymb,eqsecnum]{article}
\pdfoutput=1
\usepackage{jcappub}
\usepackage{graphicx}
\usepackage{bm}
\usepackage{amsmath}
\usepackage{hyperref}
\usepackage{url}
\usepackage[latin9]{inputenc}

\def \lleq {\lower0.9ex\hbox{ $\buildrel < \over \sim$} ~}
\def \ggeq {\lower0.9ex\hbox{ $\buildrel > \over \sim$} ~}

\def \lcdm    {$\Lambda$CDM }

\def \beq  {\begin{equation}}
\def \eeq  {\end{equation}}
\def \ber  {\begin{eqnarray}}
\def \eer  {\end{eqnarray}}

\newcommand{\newc}{\newcommand}
\newc{\be}{\begin{equation}}
\newc{\ee}{\end{equation}}
\newc{\ba}{\begin{eqnarray}}
\newc{\ea}{\end{eqnarray}}
\newc{\bea}{\begin{eqnarray*}}
\newc{\eea}{\end{eqnarray*}}
\newc{\D}{\partial}
\newc{\ie}{{\it i.e.} }
\newc{\eg}{{\it e.g.} }
\newc{\etc}{{\it etc.} }
\newc{\etal}{{\it et al.}}
\newcommand{\nn}{\nonumber}
\newc{\ra}{\rightarrow}
\newc{\lra}{\leftrightarrow}
\newc{\lsim}{\buildrel{<}\over{\sim}}
\newc{\gsim}{\buildrel{>}\over{\sim}}

\title{Hints of dark energy anisotropic stress using Machine Learning}

\author{Rub\'{e}n Arjona,}
\author{Savvas Nesseris}
\affiliation{Instituto de F\'isica Te\'orica UAM-CSIC, Universidad Auton\'oma de Madrid,
Cantoblanco, 28049 Madrid, Spain}
\emailAdd{ruben.arjona@uam.es}
\emailAdd{savvas.nesseris@csic.es}

\abstract{Recent analyses of the Planck data and quasars at high redshifts have suggested possible deviations from the flat $\Lambda$ cold dark matter model ($\Lambda$CDM), where $\Lambda$ is the cosmological constant. Here we use machine learning methods to investigate any possible deviations from $\Lambda$CDM at both low and high redshifts by using the latest cosmological data. Specifically, we apply the Genetic Algorithms to explore the nature of dark energy (DE) in a model independent fashion by reconstructing its equation of state $w(z)$, the growth index of matter density perturbations $\gamma(z)$, the linear DE anisotropic stress $\eta_\textrm{DE}(z)$ and the adiabatic sound speed $c_\textrm{s,DE}^2(z)$ of DE perturbations. We find a $\sim2\sigma$ deviation of $w(z)$ from -1 at high redshifts, the adiabatic sound speed is negative at the $\sim2.5\sigma$ level at $z=0.1$ and a $\sim2\sigma$ deviation of the anisotropic stress from unity at low redshifts and $\sim4 \sigma$ at high redshifts. These results hint towards either the presence of an non-adiabatic component in the DE sound speed or the presence of DE anisotropic stress, thus hinting at possible deviations from the $\Lambda$CDM model.}

\begin{document}
\maketitle
\flushbottom


\section{Introduction}
Through the observations of distant Type Ia supernovae at the turn of the previous century, it was discovered that the Universe is undergoing a phase of accelerated expansion on very large scales, apparently caused by a repulsive force, usually attributed to the cosmological constant \cite{Riess:1998cb,Perlmutter:1998np}. Further observations and theoretical developments led to a unified description for the formation and evolution of the Universe within the framework of General Relativity (GR), known as the standard $\Lambda$ cold dark matter model ($\Lambda$CDM), that contains only six free parameters describing the matter and dark energy (DE) content of the Universe. This model is so far the best phenomenological fit to the data \cite{Aghanim:2018eyx}.

While this framework is very successful, there also remain some tensions to resolve, such as the nature of the dominant cold dark matter component or the hotly debated Hubble constant tension, where the determination of $H_0$ deduced from physics of the early universe, i.e. the Cosmic Microwave Background (CMB) observations \cite{Aghanim:2018eyx}, is lower that the local determination of $H_0$ based on Cepheid variable-calibrated Type Ia supernovae (SnIa) \cite{Riess:2019cxk} at the $4.4\sigma$ confidence level. Also, as mentioned in Ref.~\cite{Benetti:2019gmo}, results from other surveys such as DES \cite{Abbott:2017smn}, the SPT Collaboration \cite{Henning:2017nuy} or the H0LiCOW collaboration \cite{Bonvin:2016crt}, with no common observational systematics between them, add to the idea that the tension is due more to the physics of the cosmological setting rather than experimental systematics, see Refs.~\cite{Riess:2020sih,Lemos:2018smw} and references there in for a recent discussion. However, recent re-analyses have shown that the main bulk of the  tension which remains between the CMB and local universe measurements is mainly due to the Cepheids \cite{Efstathiou:2020wxn}.

However, recently other tensions have also appeared. Using quasars in high redshifts up to $z\simeq 7.5$, it was shown in Ref.~\cite{Risaliti:2018reu} that a $\sim4\sigma$ deviation from the \lcdm model exists, suggesting a time evolution of the DE equation of state at high redshifts. On the other hand, in Refs.~\cite{Handley:2019tkm,DiValentino:2019qzk} it was shown that an enhanced lensing amplitude still present in the Planck 2018 CMB data, can be explained by a positive curvature Universe, thus violating the underlying assumptions of the flat \lcdm model.

These issues have motivated several analyses trying to reassess the level of tensions and deviations from the \lcdm model \cite{Velten:2019vwo,DiValentino:2019jae,Lin:2019htv,Camarena:2019moy,Handley:2019wlz,Gariazzo:2018zho,Guo:2018ans} or to resolve it with new physics \cite{Clifton:2011jh}. The latter approach postulates that GR is only accurate on small scales and modifications at larger scales are needed. One side-effect of this deviation from GR is that the Newtonian potentials $\Phi$ and $\Psi$ are now in general not equal, thus resulting to an anisotropic stress which could be detected from weak-lensing \cite{Huterer:2010hw}. The anisotropic stress is usually modeled via the parameter $\eta_\textrm{DE}\equiv \frac{\Phi}{\Psi}$, where $\Phi$ and $\Psi$ are the Newtonian potentials, taken to be equal in GR in the absence of anisotropic stresses from other sources such as neutrinos. Thus, any deviation of $\eta_\textrm{DE}$ from unity would point to modified gravity or if neglected, it could bias the cosmological parameters inferred from the data \cite{Cardona:2019qaz}.

If this modification of gravity is interpreted via the effective fluid approach \cite{Arjona:2018jhh,Arjona:2019rfn}, then the presence of anisotropic stress also implies that the sound speed of propagation of the DE perturbations $c_\textrm{s,DE}^2$ can be negative. However, the perturbations can still remain stable if the effective sound speed, defined as the sum of the DE sound speed and the anisotropic stress, is positive \cite{Cardona:2014iba}. Therefore, if direct measurements of $c_\textrm{s,DE}^2$ find that it is negative, this would be a smoking gun signature for the existence of an anisotropic stress and possible modifications of gravity. Furthermore, it has been shown that the effects of the anisotropic stress can be mimicked by a varying adiabatic sound speed of DE perturbations \cite{Koivisto:2005mm,Mota:2007sz}. A related quantity is also the $F(z)$ test of Ref.~\cite{Busti:2015aqa}, which is proportional to the DE sound speed and is supposed to be equal to zero for the \lcdm model. As both $F(z)$ and $c_\textrm{s,DE}^2$ are related, here we will only consider the latter.

The large scale structure (LSS) of the Universe provides a natural testbed to search for deviations from GR, since it is very sensitive to the underlying gravitational theory which directly affects the evolution of matter density perturbations. In linear theory these are parameterized via the growth parameter $\delta_m=\frac{\delta \rho_m}{\bar{\rho}_m}$ and its logarithmic derivative $f\equiv \frac{d \ln \delta_m}{d\ln a}$ called the growth-rate, where $\bar{\rho}_m$ is the background matter density and $\delta \rho_m$ its perturbation to linear order. The growth-rate can also be expressed in terms of the $\gamma$ parameter, which is useful when looking for deviations from GR, as in the \lcdm model $\gamma\simeq 6/11$ and is defined via $f(z)=\Omega^{\gamma}_m(z)$. In the \lcdm model the fact that the growth rate is scale-invariant on large scales makes it a key discriminator \cite{Franco:2019wbj}.

The main advantage of the growth is that over time it can provide information about gravity and DE and how both can be evolving as the Universe expands. The reason for this is that LSS observations in cosmology have the advantage of requiring only linear physics, which makes them an especially clean and highly successful probe \cite{Akrami:2018vks}. They can help in understanding what is the expansion rate of the Universe and how do structures form within the cosmological background. At the perturbations level, the growth of matter perturbations provides a useful tool to investigate the matter distribution in the Universe, and, more importantly, it can be measured from observations. The measurement of the growth index provides an efficient way to discriminate between modified gravity models and DE models which are developed in the context of GR \cite{Basilakos:2017rgc}. The effect of DE on the growth of perturbations is therefore an important tool in discriminating models from \lcdm \cite{Luna:2018tot} and models that are fully degenerate at the background level \cite{Nesseris:2013fca,Arjona:2018jhh,delaCruzDombriz:2006fj,Multamaki:2005zs,Pogosian:2007sw}.

Cosmology has now reached a level of precision allowing it to become a complementary probe of particle and fundamental physics. Observations of current and future surveys such as LSST \cite{Abell:2009aa}, DES \cite{Abbott:2017wau}, eBOSS \cite{Blanton:2017qot}, J-PAS \cite{Benitez:2014ibt}, DESI \cite{Aghamousa:2016zmz}, SKA \cite{Carilli:2004nx} and $21$-cm data \cite{Trott:2019lap} will allow us to probe the whole epoch from recombination to now and provide a vast amount of data for a broad span of redshifts with hundreds of thousands of supernovas type Ia, along with millions of galaxies and quasars. Clearly, the acquisition of such vast amounts of data means that traditional statistical inference is impractical, as the dimensionality of the data will also increase exponentially, a phenomenon known as ``curse of dimensionality" \cite{bellman1961adaptive}. This makes it an excellent testing ground for machine learning (ML) methods as the latter are ideal in cases where traditional fitting methods give poor results or completely fail, such as in the case of big data, but also when the parameter space is very large, too complex or not well enough understood, as is the case of DE.

As a result, machine learning will play a big role in testing accurately the standard model of cosmology, but will also help in the search for new physics and tensions in the data by placing tighter constraints on cosmological parameters \cite{Huerta:2019rtg,Arjona:2020doi,Benisty:2020kdt}. While a central goal of modern ML research is to learn and extract important features directly from data \cite{mehta2014exact}, ML methods have also been applied to reconstruct null tests of $\Lambda$CDM, i.e. quantities that are supposed to be exactly constant for all redshifts \cite{Arjona:2019fwb,Nesseris:2010ep,Nesseris:2014qca,Sapone:2014nna,Rajpaul:2012wu,Montiel:2014fpa,Yahya:2013xma}. In this paper we present a unified ML analysis of all the currently available cosmological data in order to reconstruct several key background and perturbations variables in a model-independent manner in order to explore the nature of DE. For example, such variables include the DE equation of state and the DE anisotropic stress, which we then use to test for deviations from $\Lambda$CDM.

\begin{figure*}[!t]
\centering
\includegraphics[width = 0.9\textwidth]{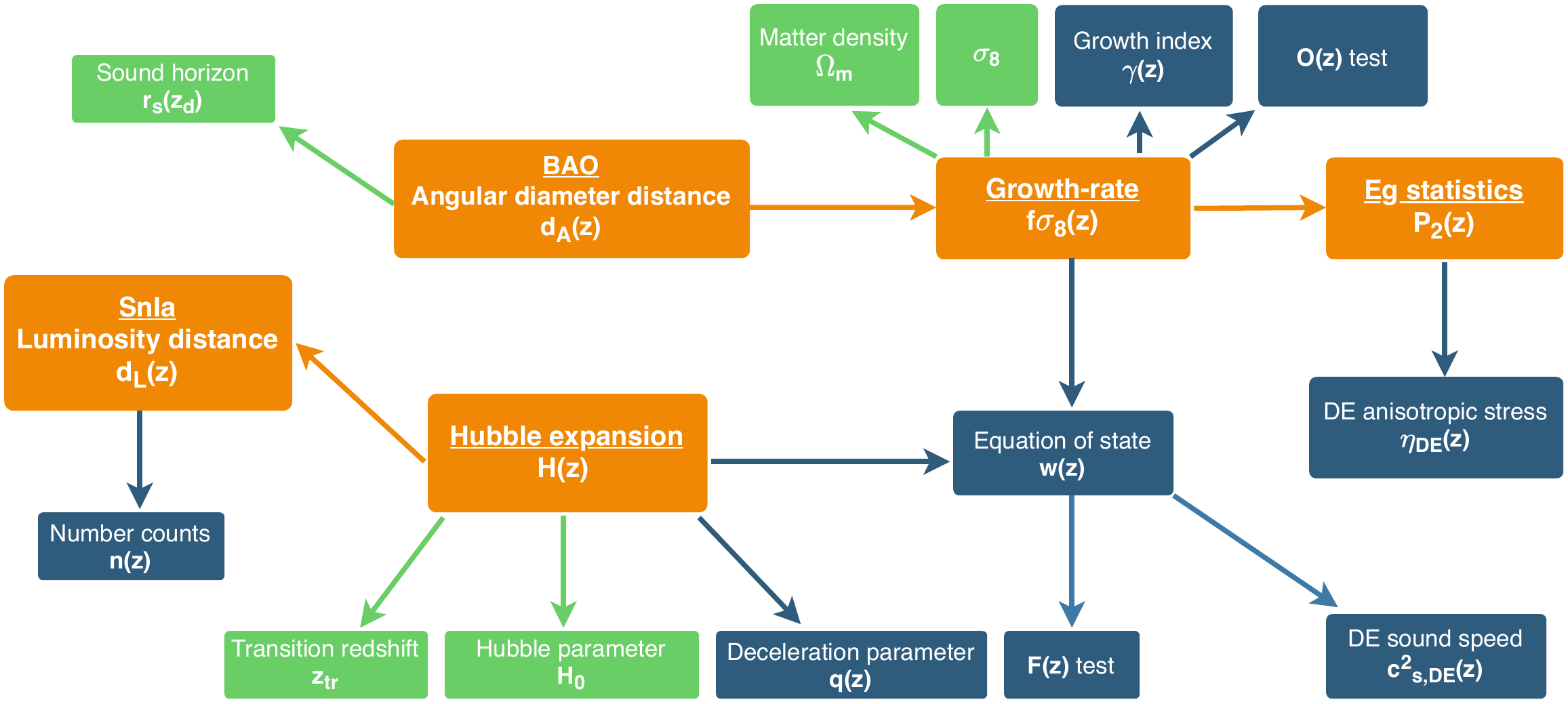}
\caption{A flowchart of the fitting process using the machine learning approach. The orange blocks represent primary quantities reconstructed directly from the data, i.e. the ML best-fits, the blue blocks represent quantities reconstructed from the latter, such as $w(z)$, and the green blocks stand for the derived (secondary) parameters. The flowchart shows the interplay between the different data and derived parameters, something which is reflected in our analysis. \label{fig:flowchart}}
\end{figure*}

The structure of our paper is as follows: In Section~\ref{sec:apptheory} we present the theoretical background of our work and our notation, in Section~\ref{sec:data} we describe in detail the cosmological data we used in our analysis, in Section~\ref{sec:ga} we present the details of our implementation of the ML approach, while in Section~\ref{sec:analysis} we present our methodology and the minimal assumptions we made for the reconstruction of the data. Then, in Section~\ref{sec:results0} we present the exact fits of the reconstruction and our main results, while in Section \ref{sec:conclusions} we present our Conclusions.

\section{Theory \label{sec:apptheory}}
Here we present some theoretical material regarding the quantities we reconstruct in what follows. The DE equation of state $w(z)\equiv \frac{P}{\rho}$ and the deceleration parameter $q(z)\equiv -\frac{\ddot{a}}{a H(a)^2}$ can be written as \cite{Bogdanos:2010yac}
\ba
w(z)&=&-1+\frac{1}{3}\left(1+z\right)\frac{d\ln \left(\Omega_\textrm{DE}(z)\right)}{d z}, \label{eq:wz} \\
q(z)&=&-1+\left(1+z\right)\frac{d\ln \left(H(z)\right)}{d z},
\ea
where $\Omega_\textrm{DE}(z)\equiv H(z)^2/H^2_0-\Omega_{m,0}(1+z)^3$ is the DE energy density. When $w(z)=-1$ we recover the \lcdm model, for which $q_0=q(z=0)=-1+\frac{3\Omega_{m0}}{2}$. Here and in what follows we will neglect radiation as it is negligible at late times when we perform our reconstructions. We also constrain the DE adiabatic sound speed $c^2_\textrm{s,DE}(z)$, which can be written in terms of the DE equation of state $w(z)$ as
\ba
c^2_\textrm{s,DE}(z)&=& \frac{\delta P_\textrm{DE}}{\delta \rho_\textrm{DE}}\nn\\
&\simeq& w(z)+\frac{1+z}{3} \frac{w'(z)}{1+w(z)}.\label{eq:cs2de}
\ea
We also consider the number counts of luminous sources, which are given by \cite{Weinberg:2008zzc}
\be
n(z)=\frac{4\pi \mathcal{N}_0 d_\textrm{L}(z)^2}{H(z) (1+z)^2}, \label{eq:numcounts}
\ee
where $\mathcal{N}_0\equiv \int_0^\infty \mathcal{N}_0(L) dL$ is the total number of sources per proper volume integrated over all luminosities. Next we also present the variables related to the matter density perturbations, in particular the growth index $\gamma(a)$, which is defined via \cite{Wang:1998gt}
\be
f(a)=\Omega_\textrm{m}(a)^{\gamma(a)},
\ee
where $f(a)=\frac{d \ln\delta_m}{d \ln a}$ is the logarithmic derivative of the growth of matter perturbations $\delta_m(a)\equiv \frac{\delta\rho_m}{\rho_m}$, the matter density is given by $\Omega_\textrm{m}(a)=\frac{\Omega_\textrm{m,0}a^{-3}}{H(a)^2/H_0^2}$ and $H(a)\equiv\frac{\dot{a}}{a}$, is the Hubble parameter as a function of the dimensionless scale factor $a=\frac{1}{1+z}$ that describes the expansion of the universe. Solving for the growth index we find that it can be expressed as
\ba
\gamma(a)&=&\frac{\ln\left(f(a)\right)}{\ln\left(\Omega_\textrm{m}(a)\right)}\nn \\
&=&\frac{\ln\left(f(a)\right)}{\ln\left(\frac{\Omega_\textrm{m,0}a^{-3}}{H(a)^2/H_0^2}\right)}.\label{eq:gamma1}
\ea
We can now proceed to reexpress the various quantities contained in Eq.~(\ref{eq:gamma1}) with ones that can be reconstructed directly from the data. Assuming a homogeneous and isotropic universe in GR, with no DE perturbations and neglecting neutrinos, then the growth factor $\delta_m(a)$ satisfies the differential equation:
\be
\delta_m''(a)+\left(\frac3{a}+\frac{H'(a)}{H(a)}\right)\delta_m'(a)-\frac32 \frac{\Omega_\textrm{m,0}}{a^5 H(a)^2/H_0^2} \delta_m(a)=0. \label{eq:ODE1}
\ee
At this point it should be noted that while here we neglect neutrinos in order to streamline the analysis, in general they can have a large effect on the amplitude and slope at LSS scales. Regarding the growth, what is measurable is not exactly the growth $\delta_m(a)$, but the combination
\ba
f\sigma_8(a)&\equiv& f(a)\cdot \sigma(a)\nn\\
&=&\frac{\sigma_8}{\delta_m(1)}~a~\delta_m'(a) ,\label{eq:fs8}
\ea
where $f(a)$ is the growth rate and $\sigma(a)=\sigma_8\frac{\delta_m(a)}{\delta_m(1)}$  is  the redshift-dependent rms   fluctuations of the linear density field at $R=8 h^{-1} \textrm{Mpc}$ while the parameter $\sigma_8$ is its value today. The combination of $f\sigma_8(a)$ is bias-free as both $f(a)$ and $\sigma_8(a)$ have a dependence on bias which is the inverse of the other, thus cancels out, and it has been shown to be a good discriminator of DE models \cite{Song:2008qt}. Performing direct manipulations of the definition of $f\sigma_8$ of Eq.~(\ref{eq:ODE1}) and Eq.~(\ref{eq:fs8}) one can show, see also Ref.~\cite{Nesseris:2011pc}, that
\ba
\hspace{-0.5cm}\frac{\delta_m(a)}{\delta_m(1)}&=&\frac{1}{\sigma_8}\int_0^a \frac{f\sigma_8(x)}{x} dx \label{eq:delta} \\
\hspace{-0.5cm}H(a)^2/H_0^2&=&\frac{3\Omega_{m,0}}{a^4 f\sigma_8(a)^2}\int_0^a dx f\sigma_8(x) \int_0^x dy \frac{f\sigma_8(y)}{y},\label{eq:H}
\ea
but also the useful relations:
\ba
\hspace{-0.5cm}\sigma_8&=&\int_0^1 \frac{f\sigma_8(x)}{x} dx \label{eq:sigma8} \\
\hspace{-0.5cm}\Omega_\textrm{m,0}&=&\frac{1}{3\int_0^1 dx \frac{f\sigma_8(x)}{f\sigma_8(1)} \int_0^x dy \frac1{y} \frac{f\sigma_8(y)}{f\sigma_8(1)}}\label{eq:Om}.
\ea
Combining Eqs.~(\ref{eq:gamma1}) and (\ref{eq:delta})-(\ref{eq:H}), we obtain our main result for the growth index:
\be
\gamma(a)=\frac{\ln\left(\frac{f\sigma_8(a)}{\int_0^a\frac{f\sigma_8(x)}{x} dx}\right)}{\ln\left(\frac{a f\sigma_8(a)^2}{3\int_0^a dx f\sigma_8(x) \int_0^x dy \frac{f\sigma_8(y)}{y}}\right)}.\label{eq:gamma2}
\ee
The main advantages of Eq.~(\ref{eq:gamma2}) are that it only requires knowledge of $f\sigma_8(a)$ and does not depend on $\Omega_{m0}$ or $H(a)$, $\sigma_8$ or any other parameter. Finally, exploiting the Noether symmetries of Eq.~\eqref{eq:ODE1} we can define a conserved charge that has to be constant at all times and redshifts, thus is an ideal null test. Following this procedure, in Ref.~\cite{Nesseris:2014mfa} it was shown that a null test for the growth can be written as
\ba
\mathcal{O}(z)&=& a^2 E(a)\frac{f\sigma_8(a)}{f\sigma_8(1)}e^{I(z)}, \\
I(z)&=&-\frac{3}{2}\Omega_{m_0}\int_{1}^a\frac{\sigma_{8_0}+\int_{1}^x\frac{f\sigma_8(y)}{y}dy}{x^4E(x)^2f\sigma_8(x)}dx,~~
\label{eq:nullf}
\ea
where we define $\sigma_{8_0}\equiv\sigma_8(a=1)$ for simplicity and we have set $E(a)\equiv \frac{H(a)}{H_0}$. It is clear that Eq.~(\ref{eq:nullf}) has to be constant for all redshifts $z$ and moreover, $\mathcal{O}(z)$ has to be equal to $1$ as
any deviation from unity might hint towards a deviation from the FLRW metric, non zero DE perturbations, a deviation from GR or a tension between the $H(z)$ and $f\sigma_8$ data.

\section{The data \label{sec:data}}
\subsection{The $H(z)$ data}
The Hubble expansion data are obtained in two complementary ways: by the clustering of galaxies or quasars and by the differential age method. The latter is connected to the redshift drift of distant objects over long time periods, usually a decade or longer, as in metric theories assuming the Friedmann-Robertson-Walker metric, the Hubble parameter can also be expressed in terms of the time derivative of the redshift as $H(z)=-\frac{1}{1+z}\frac{dz}{dt}$ \cite{Jimenez:2001gg}. The former approach is connected to the clustering of quasars or galaxies and is a direct probe of the Hubble expansion by determining the BAO peak in the radial direction \cite{Gaztanaga:2008xz}. In this analysis we use the $36$ points of the compilation from Ref.~\cite{Arjona:2018jhh}, which spans over a redshift range of $0.07\le z \le 2.34$. The data are in the form $(z_i,H_i,\sigma_{H_i})$ and are explicitly shown in Table~\ref{tab:Hzdata}. We can then minimize the $\chi^2$ analytically over $H_0$ and the result is
\ba
\chi^2_\textrm{H}&=&A-\frac{B^2}{\Gamma},\label{eq:chi2H}\\
H_0&=&\frac{B}{\Gamma},\label{eq:H0bf}
\ea
where the parameters $A$, $B$ and $\Gamma$ are defined as
\ba
A&=&\sum_i^{N_\textrm{H}}\left(\frac{H_i}{\sigma_{H_i}}\right)^2, \\
B&=&\sum_i^{N_\textrm{H}}\frac{H_i~E^\textrm{th}(z_i)}{\sigma_{H_i}^2}, \\ \Gamma&=&\sum_i^{N_\textrm{H}}\left(\frac{E^\textrm{th}(z_i)}{\sigma_{H_i}}\right)^2,
\ea
where we denote the theoretical value of the Hubble parameter as $E^\textrm{th}(z)=H^\textrm{th}(z)/H_0$ and we set $N_\textrm{H}=36$. The aforementioned data can be used to measure the Hubble constant $H_0$, to determine the deceleration transition redshift, to constrain the spatial curvature of the Universe along with distance redshift data, but also constrain the non-relativistic matter and DE parameters, as shown in \cite{Yu:2017iju}. The Hubble constant $H_0$ has been the focus of an extended discussion in the literature, in light of a tension between local and high-redshift measurements of the parameter, see Ref.~\cite{Riess:2020sih,Lemos:2018smw} and references there in for a recent discussion.

\begin{table}[!t]
\caption{The $H(z)$ data used in our analysis (in units of $\textrm{km}~\textrm{s}^{-1} \textrm{Mpc}^{-1}$). This compilation, which was presented in Ref.~\cite{Arjona:2018jhh}, is partly based on those of Refs.~\cite{Moresco:2016mzx} and \cite{Guo:2015gpa}.\label{tab:Hzdata}}
\centering
\begin{tabular}{cccccccccc}
\\
\hline\hline
$z$  & $H(z)$ & $\sigma_{H}$ & Ref.   \\
\hline
$0.07$    & $69.0$   & $19.6$  & \cite{Zhang:2012mp}  \\
$0.09$    & $69.0$   & $12.0$  & \cite{STERN:2009EP} \\
$0.12$    & $68.6$   & $26.2$  & \cite{Zhang:2012mp}  \\
$0.17$    & $83.0$   & $8.0$   & \cite{STERN:2009EP}    \\
$0.179$   & $75.0$   & $4.0$   & \cite{MORESCO:2012JH}   \\
$0.199$   & $75.0$   & $5.0$   & \cite{MORESCO:2012JH}   \\
$0.2$     & $72.9$   & $29.6$  & \cite{Zhang:2012mp}   \\
$0.27$    & $77.0$   & $14.0$  & \cite{STERN:2009EP}   \\
$0.28$    & $88.8$   & $36.6$  & \cite{Zhang:2012mp}  \\
$0.35$    & $82.7$   & $8.4$   & \cite{Chuang:2012qt}   \\
$0.352$   & $83.0$   & $14.0$  & \cite{MORESCO:2012JH}   \\
$0.3802$  & $83.0$   & $13.5$  & \cite{Moresco:2016mzx}   \\
$0.4$     & $95.0$   & $17.0$  & \cite{STERN:2009EP}    \\
$0.4004$  & $77.0$   & $10.2$  & \cite{Moresco:2016mzx}   \\
$0.4247$  & $87.1$   & $11.2$  & \cite{Moresco:2016mzx}   \\
$0.44$    & $82.6$   & $7.8$   & \cite{Blake:2012pj}   \\
$0.44497$ & $92.8$   & $12.9$  & \cite{Moresco:2016mzx}   \\
$0.4783$  & $80.9$   & $9.0$   & \cite{Moresco:2016mzx}   \\
\hline\hline
\end{tabular}~~~~~~~~
\begin{tabular}{cccccccccc}
\\
\hline\hline
$z$  & $H(z)$ & $\sigma_{H}$ & Ref.   \\
\hline
$0.48$    & $97.0$   & $62.0$  & \cite{STERN:2009EP}   \\
$0.57$    & $96.8$   & $3.4$   & \cite{Anderson:2013zyy}   \\
$0.593$   & $104.0$  & $13.0$  & \cite{MORESCO:2012JH}  \\
$0.60$    & $87.9$   & $6.1$   & \cite{Blake:2012pj}   \\
$0.68$    & $92.0$   & $8.0$   & \cite{MORESCO:2012JH}    \\
$0.73$    & $97.3$   & $7.0$   & \cite{Blake:2012pj}   \\
$0.781$   & $105.0$  & $12.0$  & \cite{MORESCO:2012JH} \\
$0.875$   & $125.0$  & $17.0$  & \cite{MORESCO:2012JH} \\
$0.88$    & $90.0$   & $40.0$  & \cite{STERN:2009EP}   \\
$0.9$     & $117.0$  & $23.0$  & \cite{STERN:2009EP}   \\
$1.037$   & $154.0$  & $20.0$  & \cite{MORESCO:2012JH} \\
$1.3$     & $168.0$  & $17.0$  & \cite{STERN:2009EP}   \\
$1.363$   & $160.0$  & $33.6$  & \cite{Moresco:2015cya}  \\
$1.43$    & $177.0$  & $18.0$  & \cite{STERN:2009EP}   \\
$1.53$    & $140.0$  & $14.0$  & \cite{STERN:2009EP}  \\
$1.75$    & $202.0$  & $40.0$  & \cite{STERN:2009EP}  \\
$1.965$   & $186.5$  & $50.4$  & \cite{Moresco:2015cya}  \\
$2.34$    & $222.0$  & $7.0$   & \cite{Delubac:2014aqe}   \\
\hline\hline
\end{tabular}
\end{table}

\subsection{The supernovae type Ia data}
We use the Pantheon SnIa compilation of Ref.~\cite{Scolnic:2017caz} of 1048 Supernovae Ia points in the redshift range $0.01<z<2.26$ along with their covariances. The apparent magnitude $m_B$ is
\be
m_\textrm{B}=5 \log_{10}\left(\frac{d_\textrm{L}(z)}{1 \textrm{Mpc}}\right)+25+M_B,
\ee
where $d_\textrm{L}(z)$ is the luminosity distance and $M_B$ the absolute magnitude. We then marginalize over the nuisance parameter $M_B$, as shown in the Appendix C of Ref.~\cite{Conley:2011ku}. Then our final expression for the $\chi^2$ is
\be
\chi^2_\textrm{SnIa}=A-\frac{B^2}{E}+\ln\left(\frac{E}{2\pi}\right),
\ee
where $A=\Delta \vec{m} \cdot \textbf{C}^{-1} \cdot \Delta \vec{m}$, $B=\Delta \vec{m} \cdot \textbf{C}^{-1} \cdot \Delta \vec{I}$ and $E=\vec{I} \cdot \textbf{C}^{-1} \cdot \vec{I}$, while $\textbf{C}$ is the SnIa covariance matrix, $\vec{I}=(1,1,\cdots,1)$ and $\Delta m\equiv m_\textrm{B,i}-m_\textrm{th}(z_i)$.

\subsection{The baryon acoustic oscillations data}
We use the BAO data from 6dFGS \cite{Beutler:2011hx}, SDDS \cite{Anderson:2013zyy}, BOSS CMASS \cite{Xu:2012hg}, WiggleZ \cite{Blake:2012pj}, MGS \cite{Ross:2014qpa} and BOSS DR12 \cite{Gil-Marin:2015nqa}, DES \cite{Abbott:2017wcz}, Lya \cite{Blomqvist:2019rah}, DR - 14 LRG \cite{Bautista:2017wwp} and quasars \cite{Ata:2017dya}. To facilitate the description of the data, we define the following functions:
\be
d_z\equiv \frac{r_\textrm{s}(z_d)}{D_\textrm{V}(z)},\label{eq:dz}
\ee
where the comoving sound speed is $r_\textrm{s}(z_d)=\int_{z_d}^\infty \frac{c_\textrm{s}(z)}{H(z)} dz$ and $z_\textrm{d}$ is the redshift at the dragging epoch, see Eq. (4) of Hu \cite{Eisenstein:1997ik}. In the \lcdm model the sound horizon can be approximated by
\be
r_\textrm{s}(z_\textrm{d})\simeq \frac{44.5\log\left(\frac {9.83} {\Omega_\textrm{m,0}h^2}\right)}{\sqrt {1+10(\Omega_\textrm{b,0}h^2)^{3/4}}} \textrm{Mpc}.\label{eq:rd}
\ee
We also define the following functions:
\be
D_\textrm{V}(z)=\left[(1+z)^2 d_\textrm{A}(z)^2 \frac{c z}{H(z)}\right]^{1/3}.
\ee
and
\be
D_\textrm{H}(z)=\frac{c}{H(z)}.
\ee
Then the 6dFGs and WiggleZ BAO data are given by
\be
\begin{array}{ccc}
 z  & d_\textrm{z} & \sigma_{d_\textrm{z} } \\
 \hline
 0.106 & 0.336 & 0.015 \\
 0.44 & 0.073 & 0.031 \\
 0.6 & 0.0726 & 0.0164 \\
 0.73 & 0.0592 & 0.0185 \\
\end{array}
\ee
where their inverse covariance matrix is
\be C_{ij}^{-1}=\left(
\begin{array}{cccc}
 \frac{1}{0.015^2} & 0 & 0 & 0 \\
 0 & 1040.3 & -807.5 & 336.8 \\
 0 & -807.5 & 3720.3 & -1551.9 \\
 0 & 336.8 & -1551.9 & 2914.9 \\
\end{array}
\right)\ee
with the $\chi^2$ given by
\be
\chi^2_\textrm{6dFS,Wig}=V^i C_{ij}^{-1} V^j,
\ee
and $V^i=d_\textrm{z,i}-d_\textrm{z}(z_i,\Omega_\textrm{m,0})$.

The BAO measurements for MGS and SDSS (LowZ and CMASS) are given by $D_\textrm{V}/r_\textrm{s} = 1/d_\textrm{z}$ via
\be\begin{array}{ccc}
 z  & 1/d_\textrm{z} & \sigma_{1/d_\textrm{z} } \\
 \hline
 0.15 & 4.46567 & 0.168135 \\
 0.32 & 8.62 & 0.15 \\
 0.57 & 13.7 & 0.12 \\
\end{array}\ee
and the $\chi^2$ as
\be
\chi^2_\textrm{MGS,SDSS}=\sum \left(\frac{1/d_\textrm{z,i}-1/d_\textrm{z}(z_i,\Omega_\textrm{m,0})}{\sigma_{1/d_\textrm{z,i}}}\right)^2.
\ee
The BAO data from DES are of the form $d_\textrm{A}(z)/r_\textrm{s}$ with $(z,d_\textrm{A}(z)/r_\textrm{s},\sigma_i)=(0.81, 10.75, 0.43)$ and the $\chi^2$ is given by \be
\chi^2_\textrm{DES}=\sum \left(\frac{d_\textrm{A}(z,i)/r_\textrm{s}-d_\textrm{A}(z_i,\Omega_\textrm{m,0})/r_\textrm{s}}{\sigma_{d_\textrm{A}(z,i)/r_\textrm{s}}}\right)^2.
\ee
while the BAO data from Lya are of the form $f_\textrm{BAO}=\bigg(\frac{(1+z)d_\textrm{A}}{r_\textrm{s}}, \frac{D_\textrm{H}}{r_\textrm{s}}\bigg)$ and are given by
\be
\begin{array}{ccc}
 z & f_\textrm{BAO}  &  \sigma_{f_\textrm{BAO}} \\
 \hline
 2.35 & (36.3,9.2) & 0.36 \\
\end{array}
\ee
with the $\chi^2$ given by
\be
\chi^2_\textrm{Lya}=\sum \left(\frac{f_\textrm{BAO,i}-f_\textrm{BAO}(z_i,\Omega_\textrm{m,0})}{\sigma_{f_\textrm{BAO}}}\right)^2.
\ee
Note that the Lya measurements are correlated, however no covariance matrix is provided in Ref.~\cite{Blomqvist:2019rah}. As we cannot include this correlation of the measurements in our analysis, this could possibly cause our errors to be underestimated.

Finally, the DR-14 LRG and quasars BAO data assume $r_\textrm{s,fid} = 147.78~\textrm{Mpc}$ and are given by $D_\textrm{V}/r_\textrm{s} = 1/d_\textrm{z}$
\be
\begin{array}{ccc}
 z  & 1/d_\textrm{z} & \sigma_{1/d_\textrm{z} } \\\hline\\
 0.72 & \frac{2353}{r_\textrm{s,fid}} & \frac{62}{r_\textrm{s,fid}} \\~\\
 1.52 & \frac{3843}{r_\textrm{s,fid}} & \frac{147}{r_\textrm{s,fid}} \\
\end{array}
\ee
and the $\chi^2$ is given by
\be
\chi^2_\textrm{LRG,Q}=\sum \left(\frac{1/d_\textrm{z,i}-1/d_\textrm{z}(z_i,\Omega_\textrm{m,0})}{\sigma_{1/d_\textrm{z,i}}}\right)^2.
\ee
The total $\chi^2$ is then
\be
\chi^2_\textrm{tot}=\chi^2_\textrm{6dFS,Wig}+ \chi^2_\textrm{MGS,SDSS} + \chi^2_\textrm{DES} +\chi^2_\textrm{Lya}+\chi^2_\textrm{LRG,Q}.
\ee
For all the aforementioned data we will use in our analysis, we assume that the likelihoods are sufficiently Gaussian so that we can add their $\chi^2$ together and just minimize the total.

Finally, all the $\chi^2$ terms in the previous equation depend on the sound horizon at the drag redshift $r_\textrm{d}=r_\textrm{s}(z_\textrm{d})$ via Eq.~\eqref{eq:dz}, which is difficult to estimate in a model independent approach. Therefore, we leave $r_\textrm{d}$ as a free parameter and obtain its value from the data by minimizing the total $\chi^2$ over it.
\subsection{The growth-rate data}
We also use the growth-rate $f\sigma_8$ compilation given in Table I of Ref.~\cite{Sagredo:2018ahx}, where the authors analyzed different subsets in the data and implemented Bayesian model comparison to test the internal robustness of the dataset. These data are obtained via the redshift-space distortions and in fact determine the combination $f\sigma_8(a)\equiv f(a)\cdot \sigma(a)$. The value of $f\sigma_8(a)$ can be directly determined from the ratio of the monopole to the quadrupole of the redshift-space power spectrum, which depends on the parameter $\beta=f/b_0$, where $b_0$ is the bias
and $f$ is the growth rate assuming linear theory \cite{Percival:2008sh,Song:2008qt,Nesseris:2006er}. It  can be shown that $f\sigma_8(a)$ is independent of the bias, as the latter completely cancels out from the previous expression.

Moreover, $f\sigma_8(a)$ has been shown to be a good discriminator of DE models \cite{Song:2008qt}. For more details on the covariance matrix of the data and how to correct for the Alcock-Paczynski effect see Refs.~\cite{Sagredo:2018ahx}, \cite{Nesseris:2017vor} and \cite{Kazantzidis:2018rnb}. The advantage of using the combination $f\sigma_8$, instead of just the growth-rate $f(z)$, is that the former is directly associated to the power spectrum of peculiar velocities of galaxies \cite{Nesseris:2014mea}.

\subsection{The $E_\textrm{g}$ data}
The flat Friedmann-Robertson-Walker (FRW) metric, which can describe accurately the geometry of the Universe, reads $ds^2=-(1+2\Psi)dt^2 +a(t)^2(1-2\Phi)dx^2$, where $a$ is the scale factor and $\Psi$ and $\Phi$ are the two scalar gravitational potentials. Then the gravitational slip, can be defined as the ratio of the gravitational potentials $\eta_\textrm{DE}=\frac{\Phi}{\Psi}$, which in GR is equal to unity. These potentials satisfy the two Poisson equations in Fourier space:
\ba
-\frac{k^2}{a^2}(\Phi+\Psi)&=&4 \pi G_\textrm{N} \Sigma(k,a) \rho_m \delta_m, \\
-\frac{k^2}{a^2}\Psi&=&4 \pi G_\textrm{N} \mu(k,a) \rho_m \delta_m,
\ea
where $G_\textrm{N}$ is the bare Newton's constant, while $\Sigma$ and $\mu$ parameterize deviations in GR. In the case of GR these parameters have the value $\Sigma=2$ and $\mu=1$.

In order to test the aforementioned relations, the $E_\textrm{g}$ statistic was created, aiming for it to be bias independent at linear order \cite{Zhang:2007nk,Reyes:2010tr}. The $E_\textrm{g}$ test can be expressed as the expectation value of the ratio of lensing and galaxy clustering observables at a scale $k$ as follows
\be
E_\textrm{g}=\left\langle \frac{a \nabla^2(\Psi+\Phi)}{3 H_0^2 f \delta_m}\right\rangle.
\ee

To derive the gravitational slip in a model independent way we reconstruct two quantities through the $E_\textrm{g}$ and $f\sigma_8$ data. The first quantity is $P_2(z)$ which is defined as $P_2=\frac{\Omega_\textrm{m,0}\Sigma}{f}$ and depends on the lensing potential and the growth rate. In GR this reduces to $P_2=\frac{2\Omega_\textrm{m,0}}{f}$, which implies that in GR we have $E_\textrm{g}=\frac{\Omega_\textrm{m,0}}{f}$. In general, $E_\textrm{g}$ can be related to the $P_2$ statistic of Ref.~\cite{Pinho:2018unz} via $P_2=2 E_\textrm{g}$. The second quantity is $P_3$, expressed as $P_3=\frac{\left(f\sigma_8(z)\right)'}{f\sigma_8(z)}$, where the prime is the derivative with respect to $\ln a$. Then, the gravitational slip can be derived in a model independent way as \cite{Pinho:2018unz}
\be
\eta_\textrm{DE}(z)=\frac{3P_2(z)(1+z)^3}{2E(z)^2\left(P_3(z)+2+\frac{E'(z)}{E(z)}\right)}-1, \label{eq:etade}
\ee
where $E(z)\equiv H(z)/H_0$. The exact data points we used are given in Table~\ref{tab:Eg} for completeness. Note that in our reconstruction we will use directly $E(z)$, see Eq.~\eqref{eq:chi2H}, so for this particular expression no value for $H_0$ is needed.

\begin{table}[!t]
\caption{The $E_\textrm{g}$ data used in this analysis as compiled by Refs.~\cite{Pinho:2018unz} and \cite{Skara:2019usd}. Note that some of the points in the previous references were duplicates as they come from the same surveys, albeit with combinations of different external probes, so we use only one of the measurements to avoid strong correlations. Here we only show the points we used in the analysis.\label{tab:Eg}}
\begin{centering}
\begin{tabular}{cccc}
 $z$ & $E_\textrm{g}$  &  $\sigma_{E_\textrm{g}}$ & \\ \hline
0.267 &	0.43 &	0.13 & \\
0.270 & 0.40 &  0.05 & \\
0.305 &	0.27 &	0.08 & \\
0.320 &	0.40 &	0.09 & \\
0.554 &	0.26 &	0.07 & \\
0.570 &	0.30 &	0.07 & \\
0.600 &	0.16 &	0.09 & \\
0.860 &	0.09 &	0.07 &
\end{tabular}
\par
\end{centering}
\end{table}

----------------------------------------------------------------------------------------------------------
\section{The Genetic Algorithms \label{sec:ga}}
\subsection{Mathematical formalism\label{sec:ga1}}
In this Section we will now discuss the background and the particular implementation of the Genetic Algorithms (GA) used in our analysis. The GAs are a particular class of stochastic optimization ML methods that specialize in unsupervised symbolic regression of data by reconstructing analytically, using one or more variables, a function that describes the data. This is achieved by mimicking the theory of evolution via the notion of natural selection, as expressed by the genetic operations of mutation and crossover. In a nutshell, a set of test functions evolves over long periods of time under the influence of the stochastic operators of crossover, i.e. the joining of two individuals to make an offspring, and mutation, i.e. a random alteration of an individual.

Mathematically, these operations can be described via an example as follows. Assuming we have two functions $f_1(x)=1+x+x^2$ and $f_2(x)=\sin(x)+\cos(x)$, then the mutation operation will stochastically, i.e. randomly, modify the coefficients, the exponents but also the various functions present in the expressions. For example, after the mutation operation is applied the functions might be changed to $f_1(x)=1+2x+x^2$ and $f_2(x)=\sin(x^2)+\cos(x)$, where in the first case the coefficient of the second term changed from one to two and in the second case, $x=x^1$ was changed to $x^2$. On the other hand the crossover operation randomly combines the two functions to produce two more, e.g. in the aforementioned example the GA might combine the terms $1+2x$ from $f_1$ and $\cos(x)$ to make $\widetilde{f}_1(x)=1+2x+\cos(x)$, while the remaining parts will combine to $\widetilde{f}_2(x)=x^2+\sin(x^2)$.

As the GA is a stochastic approach by nature, the probability that a population of functions will produce offspring is frequently assumed to be proportional to its fitness to the data, this being a $\chi^2$ statistic in our analysis. Specifically, we choose the $\mathcal{N}$ best-fitting functions, where $\mathcal{N}= sel \times pop$, $sel$ is the selection percentage (usually $sel\sim 10\%$) and $pop$ is the population size (usually $pop=100$), so that the mutation and crossover stochastic operators are applied to these chosen functions. The stochastic nature of the mutation operator in this case means that a function is changed randomly, with the various coefficients and exponents drawn from a uniform random distribution $\textbf{X}\sim U(-9,9)$, while the crossover is applied upon two uniform randomly chosen functions, from the whole set of best-fitting GA functions. For example, if there are 10 best-fitting GA functions, the GA will randomly choose for the crossover two out of the ten and proceed to apply the crossover operation as described in the previous paragraph. This is done by choosing a uniformly distributed random number in the range between one and the length of the expression, such that the two functions can be intermixed. Finally, the mutation will also be applied randomly, as described before, only if a uniformly random drawn number $\textbf{X}\sim U[0,1]$ does not exceed the mutation rate, typically set to $0.3$. For further details on the GA and some applications to cosmology see Refs.~\cite{Bogdanos:2009ib,Nesseris:2012tt}.

The reconstruction procedure then proceeds as follows. First, we select a set of analytic functions, commonly called the ``grammar", with which we set up the first generation. In this first step we also impose any necessary priors dictated either by physical or mathematical reasons. For example, we may impose that the value of the luminosity distance at $z=0$ is zero, i.e. $d_\textrm{L}(z=0)=0$ or that the Hubble parameter today is $H(z=0)=H_0$. Hence, we make no assumptions on the curvature of the Universe or on any modified gravity model.

Note that we do not only use grammars including simple polynomials, but we also include several other functions, see Table \ref{tab:grammars} for a complete list. More complicated forms like $e^{f(z)}$ and $f(z)^{g(z)}$ are then automatically created by the mutation and crossover operations as described earlier. Also, it should be stressed that the choice of the grammar and the population size has already been tested in Ref.~\cite{Bogdanos:2009ib}\footnote{See for example Fig. 2 of Ref.~\cite{Bogdanos:2009ib} for the effect of the population size on the convergence and a discussion in page 5 for the effects of the grammar.}. Similarly, the seed numbers are also crucial as they are used to create the initial population of functions used later on by the GA. This is clearly analogous to the initialization of weights in a neural network.

\begin{table}[!t]
\caption{The grammars used in the GA analysis, which include polynomials, exponentials, fractions, constants, trigonometric functions, logarithms etc. More complicated forms are automatically created by the mutation and crossover operations as described in the text.\label{tab:grammars}}
\begin{centering}
\begin{tabular}{cc}
 Grammar type & functions \\ \hline
Polynomials & $c$, $x$, $1+x$ \\
Fractions & $\frac{x}{1+x}$\\
Trigonometric & $\sin(x)$, $\cos(x)$, $\tan(x)$\\
Exponentials & $e^x$, $x^x$, $(1+x)^{1+x}$ \\
Logarithms & $\log(x)$, $\log(1+x)$
\end{tabular}
\par
\end{centering}
\end{table}

After we have constructed this initial population of functions, the fitness of each member is determined by a $\chi^2$ statistic using the data and the best-fitting functions in every generation are selected via a tournament selection, see Ref.~\cite{Bogdanos:2009ib} for more details. Afterwards, the two stochastic operations of the mutation and the crossover are consecutively applied and the whole process is repeated several thousands of times so as to ensure convergence. In an analogy to traditional Monte Carlo approaches, we also reran the GA with a plethora of different random seeds, so as to explore the functional space.

\subsection{Error analysis\label{sec:ga2}}
After the GA has converged, the final output is a set of differentiable and continuous  functions of the redshift $z$. In order to obtain an estimate of the errors in the reconstructed functions we follow the path integral approach of Refs.~\cite{Nesseris:2012tt,Nesseris:2013bia}. We approximate the errors using the path integral approach expanded, for a limited set of functions, around the best fit GA. Whilst this may not be representative of the true errors, we find they agree well with Fisher matrix and bootstrap Monte-Carlo analyses \cite{Nesseris:2012tt}.

Then, given a function $f(x)$ which is reconstructed by the GA, the ``path integral'' approach of Ref.~\cite{Nesseris:2012tt} can provide us with the $1\sigma$ error $\delta f(x)$. 
Moreover, in principle the GA may in fact consider any possible function  as it is exploring the functional space and even though possibly poor fits may be discarded, they can indeed contribute in the total probability when integrating over the likelihood and as a result, they have to be taken into account in the error calculations.

This approach has been well-tested by comparing the GA errors against Fisher and Bootstrap MCMC analyses, and in Ref.~\cite{Nesseris:2012tt} (see Fig.~2) it was demonstrated that the ``path integral" approach gives robust and reliable errors, not affected by the error reconstruction method. For the ``path integral" approach employed here, we assume the errors correspond to $1\sigma$, which is equal to one standard deviation of a normal deviation around the GA best-fit. This is similar to error propagation assuming that the error in a quantity is given by $\sigma_{ f}=f'(p) \delta p$, where $p$ in this case is a parameter. Having extensively tested the GA against Fisher matrix and Bootstrap MCMCs in Ref.~\cite{Nesseris:2012tt}, where the GA was confirmed to give robust error estimates, we find this assumption of the error propagation to be appropriate for the data used here. Then, we can treat the values of the function $f$ as random variables described by a normal distribution
\be
\mathcal{L}(f)=\frac{e^{-\frac{(f-f_\textrm{GA})^2}{2 \delta f^2}}}{\sqrt{2 \pi}  \delta f}.
\ee
This assumption is justified in the context of the ``path integral" approach, thus to good approximation the GA error can in fact be taken to be Gaussian, see also Eqs.~(2.5) and (2.6) in Ref.~\cite{Nesseris:2012tt}.

Then, the error propagated to any quantity formed by the function $f$, eg $g=g(f)$ can be estimated by using the definition of the standard deviation $\delta g^2=\langle g^2\rangle-\langle g\rangle^2$ and the expectation value $\langle g\rangle=\int_{-\infty}^{+\infty} g(f)~\mathcal{L}(f)df$. For example, we demonstrate this approach for the simple example of $g(f)=f^2$, where we would expect the error of $g$ to be $\delta g=2 f_\textrm{GA} \delta f+\cdots$. Indeed, we find
\ba
\langle g^2\rangle &=&3 \delta f^4+f_\textrm{GA}^4+6 \delta f^2 f_\textrm{GA}^2 \nn \\
\langle g\rangle^2 &=& \left(\delta f^2+f_\textrm{GA}^2\right)^2,
\ea
which gives
\ba
\delta g & = & \sqrt{2 \delta f^4+4 \delta f^2 f_\textrm{GA}^2}\nn \\
& \simeq & 2 f_\textrm{GA} \delta f+\cdots,
\ea
in agreement with the expected value. Similarly, one can derive for example the error of the $\textrm{Om}$ statistic, which can be reconstructed using the GA best-fit $H_\textrm{GA}(z)$, see Ref.~\cite{Arjona:2019fwb}. The $\textrm{Om}$ statistic is a null test, i.e. a true/false statement or equivalently a consistency test, that has to be true at all redshifts if the \lcdm is the real underlying physical model. It is derived by solving the Friedmann equation \cite{Sahni:2008xx,Nesseris:2010ep}
\be
H(z)^2 / H_0^2 =\Omega_\textrm{m,0} (1+z)^3+1-\Omega_\textrm{m,0},
\ee
for $\Omega_\textrm{m,0}$ using simple algebraic manipulations, so its value should be equal to that of $\Omega_\textrm{m,0}$ if and only if the \lcdm is the true model. Thus, using a model-independent approach to reconstruct the Hubble parameter $H(z)$ we can probe for deviations from \lcdm in a straight-forward manner. Noting that the $\textrm{Om}$ statistic is defined as
\ba
\textrm{Om}(z)&=&\frac{H(z)^2/H_0^2-1}{(1+z)^3-1},
\ea
then, using the aforementioned approach we find the error on the $\textrm{Om}$ statistic is
\ba
\delta \textrm{Om}(z)&=&\frac{2 H_\textrm{GA}(z)\delta H(z)/H_0^2}{(1+z)^3-1},
\ea
as expected for traditional error propagation as well.

Similarly, one can derive the error propagation of a quantity that depends on two reconstructed quantities by the GA. The procedure is exactly the same as before and we now consider the example of the DE energy density parameter $\Omega_\textrm{DE}=\frac{H(z)^2}{H_0^2}-\Omega_\textrm{m,0} (1+z)^3$, where we assume the two reconstructed quantities are $H(z)$ and $\Omega_\textrm{m,0}$, each being described by a normal distribution. In this case the error on $\Omega_\textrm{DE}$ can be found to be:
\ba
\delta \Omega_\textrm{DE}(z)^2 \simeq 4 \frac{H(z)^2}{H_0^2}\frac{\delta H(z)^2}{H_0^2}+(1+z)^6 \delta \Omega_\textrm{m,0}^2+\cdots,~~~
\ea
again in agreement with the expected value from standard error propagation.

We also have to calculate quantities that contain derivatives, such as the DE equation of state or the deceleration parameter. In this case we will assume that we can model the error propagation as a variation of the functions during the evolution of the functional space of the GA, i.e. $\delta f=\delta (f)$. This is in agreement with the previous approach as if we assume $g=f^2$ then we have $\delta g=2 f \delta f$ as expected. We can further assume that the variational $\delta$ commutes with derivatives, i.e. $\delta (\frac{df}{dx})=\frac{d}{dx}(\delta f(x))$. The reason for this is that we can always assume that at any point $x$ functions $f$ that are close to the best-fit, can be written as $f\simeq f_\textrm{GA}+\delta f$, so that $\frac{df}{dx}\simeq \frac{df_\textrm{GA}}{dx}+\frac{d \delta f}{dx}$, which implies $\delta (\frac{df}{dx})\simeq  \frac{df}{dx}- \frac{df_\textrm{GA}}{dx}\simeq\frac{d \delta f}{dx}$ as mentioned before. For example, in the case of the deceleration parameter we have:
\ba
q_\textrm{GA}(z)&=&-1+(1+z)\frac{d \ln H_\textrm{GA}}{dz},
\ea
which implies that the error is
\ba
\delta q(z)&=&(1+z)\delta \left[\frac{d \ln H}{dz}\right] \nn\\
&=& (1+z)\frac{d }{dz}\left[\delta\ln H\right]\nn\\
&=& (1+z)\frac{d }{dz}\left[\frac{\delta H}{H_\textrm{GA}}\right].
\ea
Similarly, we will assume that the variational $\delta$ commutes with integrals, an assumption commonly made in variational calculus, so that for example for the $\sigma_8$ parameter we have:
\ba
\sigma_{8,\textrm{GA}}&=&\int_0^1 \frac{f\sigma_{8,\textrm{GA}}(x)}{x}dx
\ea
and the corresponding error on the derived GA best-fit is
\ba
\delta\sigma_8&=&\int_0^1 \frac{\delta f\sigma_8(x)}{x}dx.
\ea
Then, for the matter density parameter $\Omega_{m0}$ we also have
\ba
\Omega_{m,0}{}_\textrm{GA}&=&\frac{1}{3\int_0^1 dx \frac{f\sigma_8{}_\textrm{GA}(x)}{f\sigma_8{}_\textrm{GA}(1)} \int_0^x dy \frac1{y} \frac{f\sigma_8{}_\textrm{GA}(y)}{f\sigma_8{}_\textrm{GA}(1)}} \nn \\
&=& \frac{1}{3\int_0^1 dx F(x) \int_0^x dy \frac1{y} F(y)},
\ea
where $F(x)=\frac{f\sigma_8{}_\textrm{GA}(x)}{f\sigma_8{}_\textrm{GA}(1)}$. Then, the error is
\ba
\frac{\delta \Omega_{m,0}}{3 \Omega_{m,0}^2 }&=&\left|\int_0^1 dx \left[ \delta F(x) \int_0^x dy \frac{F(y)}{y}+F(x) \int_0^x dy \frac{\delta F(y)}{y}\right]\right|, \nn\\
\ea
where we have set
\ba
\delta F(a)&=& \delta \left(\frac{f\sigma_8(a)}{f\sigma_8(1)}\right) \nn \\
&=&\frac{\delta f\sigma_8(a)}{f\sigma_8{}_\textrm{GA}(1)}-\frac{f\sigma_8{}_\textrm{GA}(a)}{f\sigma_8{}_\textrm{GA}(1)^2}~\delta f\sigma_8(1).
\ea

\subsection{General considerations\label{sec:ga3}}
Even though we have made no assumptions on a particular DE model, such as the $\Lambda$CDM, it may happen that the data contain some model assumptions. A notorious such example is the JLA SnIa compilation \cite{Betoule:2014frx}, where the cosmological parameters have to fitted simultaneously with variables of astrophysical origin, such as the light-curve parameters. In a similar vein, in the Pantheon compilation \cite{Scolnic:2017caz}, some model dependence still remains, even though the light-curve parameters have already been marginalized over. This is due to the fact that the SnIa surveys have to include in the analysis particular peculiar velocity corrections, assuming the \lcdm model and linear theory\cite{Mohayaee:2020wxf}. Furthermore, in order to derive the covariance matrix, frequently a fiducial model is assumed \cite{Scolnic:2017caz}. As we will see later on, the best-fit is close to the \lcdm model in our analysis, hence we can safely assume that these effects do not affect the reconstruction process.

Also, there exist other non-parametric approaches such as Gaussian processes (GP). These assume the data can be described by a stochastic Gaussian process, one can later on map to any cosmological function of interest~\cite{Shafieloo:2012ht,Busti:2014aoa,Pinho:2018unz,Bengaly:2019oxx}. Even though the GP necessitate choosing both a kernel and a fiducial model for the mean value, in Ref.~\cite{Shafieloo:2012ht} it has been shown that these assumptions do not affect the results of the reconstruction. However, the GA have the advantage over the GP that they need no physical model prior assumptions, for example of a flat Universe or a DE model, other than the choice of the grammar which has been show it only affects convergence rate~\cite{Bogdanos:2009ib}. By comparing plots of the same reconstructed parameter we can see that both reconstruction methods give similar errors, see for example the plot of $H(z)$ in Fig.~2 of Ref.~\cite{Pinho:2018unz}.

Finally, we should stress that it has been demonstrated with the use of mock data sets that the GA can successfully recover the underlying physical model. For example, in the case of generic, i.e. non-physically motivated, mock data this was done in Ref.~\cite{Nesseris:2012tt} (see Fig. 2), where as it can be seen the GA matches both the underlying fiducial model and also gives error estimates that are similar with other approaches (Fisher and MCMC). Similarly, in the case of cosmological data the ability of the GA to recover the fiducial model was tested in Ref.~\cite{Martinelli:2020hud} (see Fig. 6) and Ref.~\cite{Hogg:2020ktc} (see Figs. 7-9) with a variety of mock cosmological data, while using several quite different fiducial models such as the \lcdm and modified gravity theories, but also models that violate the distance duality relation.

\section{Methodology \label{sec:analysis}}
We will now describe how to reconstruct using the GA the Hubble parameter $H(z)$ from the Hubble expansion history $H(z)$ data, the luminosity distance $d_\textrm{L}(z)$ from the Pantheon Type Ia supernovae (SnIa) data, the angular diameter distance $d_\textrm{A}(z)$ from Baryon Acoustic Oscillations (BAO), $f\sigma_8(z)$ from the growth-rate data obtained via the redshift-space distortions (RSD) and $P_2(z)$ from the $E_\textrm{g}$ data.

These functions will in turn be used to reconstruct the DE anisotropic stress $\eta_\textrm{DE}(z)$, the growth index $\gamma(z)$, the DE equation of state $w(z)$ and the DE adiabatic sound speed $c_\textrm{s,DE}^2(z)$. Furthermore, we will also reconstruct the growth rate null test $\mathcal{O}(z)$ presented in Ref.~\cite{Nesseris:2014mfa} as a consistency test of the \lcdm model and the number counts of luminous sources $n(z)$. We also derive the matter density $\Omega_\textrm{m,0}$ and the root mean square (rms) density fluctuation $\sigma_8$ from the $f\sigma_8$ data, the value of the Hubble constant $H_0$ and the sound horizon at the drag epoch $r_\textrm{d}$ from the BAO data.



As mentioned in the previous section, in order to reconstruct the data we will only make very few minimal physical or mathematical assumptions, but we will make no assumption of a DE model or that the spatial curvature of the Universe is zero, i.e. flatness. However, we will assume homogeneity, isotropy and the Friedmann-Robertson-Walker (FRW) metric. Specifically we have assumed:
\begin{enumerate}
  \item The Hubble parameter today is given by the Hubble constant $H(z=0)=H_0$. Then, $H_0$ is estimated directly from the $H(z)$ data.
  \item We assume the Hubble law at low redshifts $d_\textrm{L}(z\simeq0)\simeq\frac{c}{H_0}z$.  We use the Hubble constant $H_0$ from the $H(z)$ fit to break the degeneracies with the absolute SnIa magnitude.
  \item Similarly, at low redshifts we assume $d_\textrm{A}(z\simeq0)\simeq\frac{c}{H_0}z$ due to the Hubble law. We make no assumptions for the sound horizon at drag redshift $r_\textrm{d}$, which is minimized over. 
  \item The Universe at early times went through a phase of matter domination ($z\simeq100)$, so the linear growth behaves as $\delta_m(a)\simeq a$ at high redshifts.
\end{enumerate}

We also note that the growth rate data has a dependence on the fiducial model which can be corrected by rescaling the measurements by the ratios of $H(z)D_\textrm{A}(z)$ as it is explained in Ref.~\cite{Nesseris:2017vor}. Finally, the SnIa contain some model dependence, as one must optimize parameters in the lightcurve function simultaneously with those of the assumed model. This mainly affects the covariance matrix, which is typically inferred based on an fiducial background model, usually $\Lambda$CDM. However, since in our case the best-fit is close to \lcdm and the errors are much larger than the effects of the model-bias in the covariance, we can safely assume for now that these effects have a minimal effect on the minimization.

For illustration purposes we also present in Fig.~\ref{fig:flowchart} a flowchart of the whole fitting process. To estimate the errors on these reconstructed quantities we use the \textit{Path Integral} approach developed by Refs.~\cite{Nesseris:2012tt,Nesseris:2013bia}, where one calculates analytically a path integral over the whole functional space that can be scanned by the GA. Then this error is propagated onto the various derived quantities with the error propagation approach described in detail in Sec.~\ref{sec:ga2}.

\section{Results \label{sec:results0}}
Here we present the results of our analysis by considering all data separately. Specifically we show the reconstructions of all relevant quantities including the DE equation of state $w_\textrm{DE}(z)$, the DE adiabatic sound speed $c_\textrm{s,DE}^2(z)$, the growth index $\gamma(z)$, the $\mathcal{O}(z)$ test, the DE anisotropic stress $\eta_\textrm{DE}(z)$, the number counts of luminous sources $n(z)$ and $\sigma_8$.

First, we fit the BAO data without assuming the value of $H_0$ given by the $H(z)$ reconstruction. To perform the reconstruction of the BAO data we minimize the $\chi^2$ over the quantity $r_\textrm{sh}=r_\textrm{s}\cdot h$, where $r_\textrm{s}$ is the sound horizon at the drag redshift and $h$ is the Hubble parameter, thus avoiding any bias of the results due to a specific value of $H_0$. This reconstruction will affect the growth-rate $f\sigma_8$, and thereby the growth index $\gamma$ and the $O(z)$ test, but also the secondary parameters: $\Omega_\textrm{m,0}$ and $\sigma_8$. This in turn affects the equation of state $w(z)$ and the dark energy sound speed $c^2_\textrm{s,DE}$ since they depend on the value of $\Omega_\textrm{m,0}$, see Eq.~\eqref{eq:wz}. Finally, the DE anisotropic stress also depends on the $f\sigma_8$ reconstruction.

\begin{figure*}[!t]
\centering
\includegraphics[width = 0.49\textwidth]{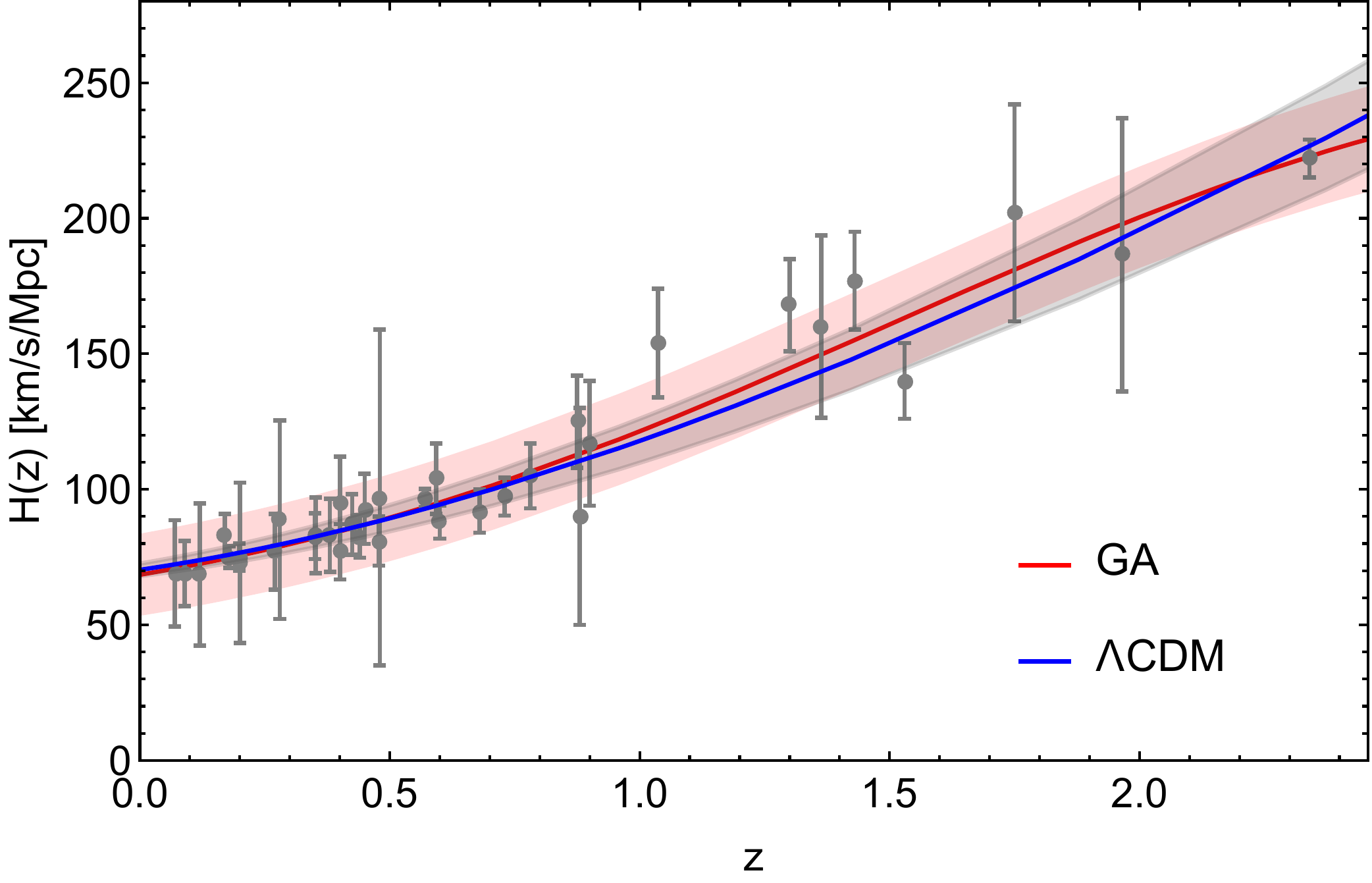}
\includegraphics[width = 0.49\textwidth]{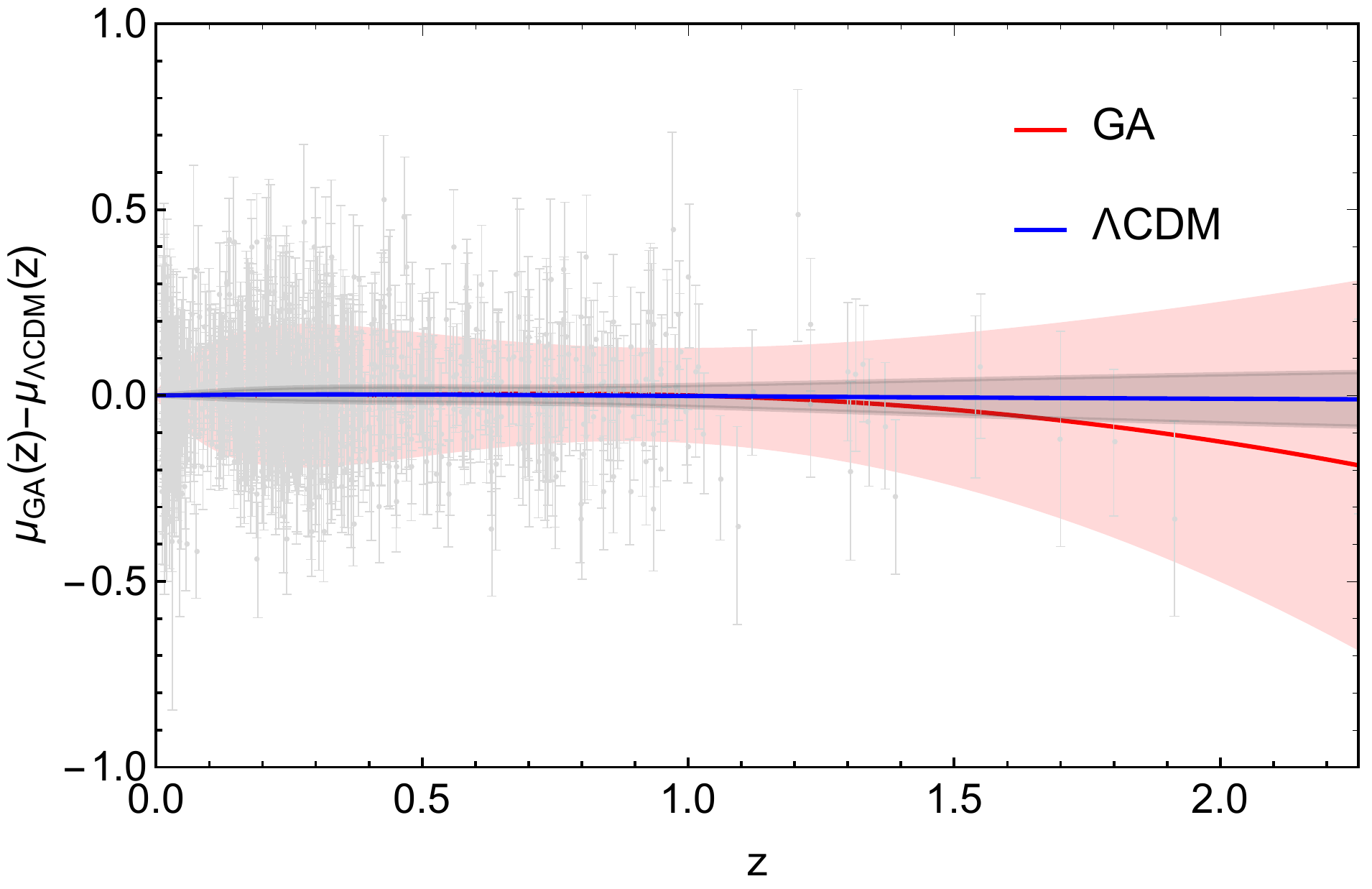}
\caption{Left: The $H(z)$ data compilation along with the \lcdm best-fit (blue line) and the GA best-fit (red line). Right: The difference between the GA best-fit distance modulus of the Pantheon SnIa data (red line) and that of the \lcdm
model (blue line). In both cases the red and grey shaded regions correspond to the $1\sigma$ confidence region for the GA and \lcdm respectively, and the actual data are shown as grey points in the background. \label{fig:Hz_snia}}
\end{figure*}

\begin{figure*}[!t]
\centering
\includegraphics[width = 0.49\textwidth]{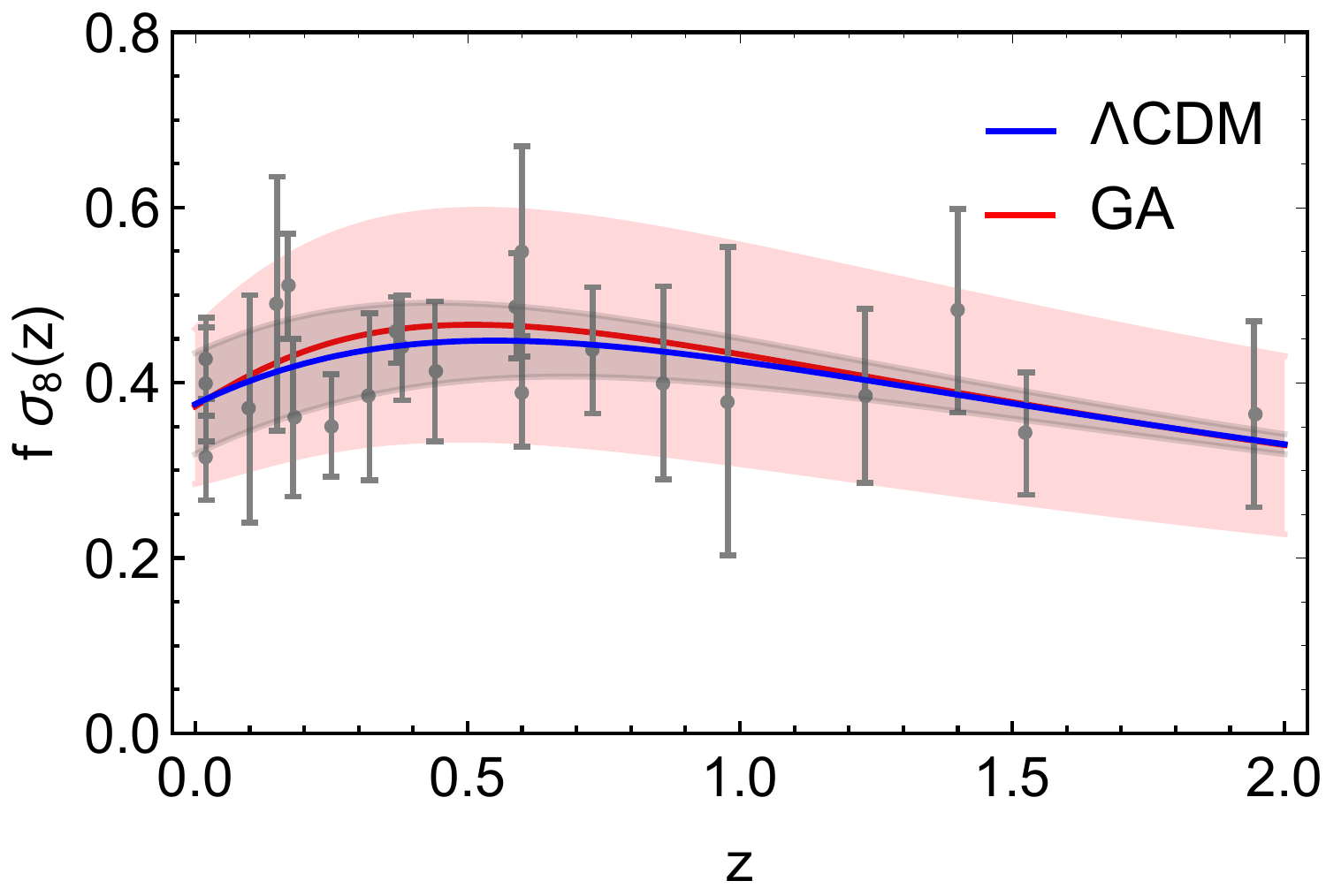}
\includegraphics[width = 0.49\textwidth]{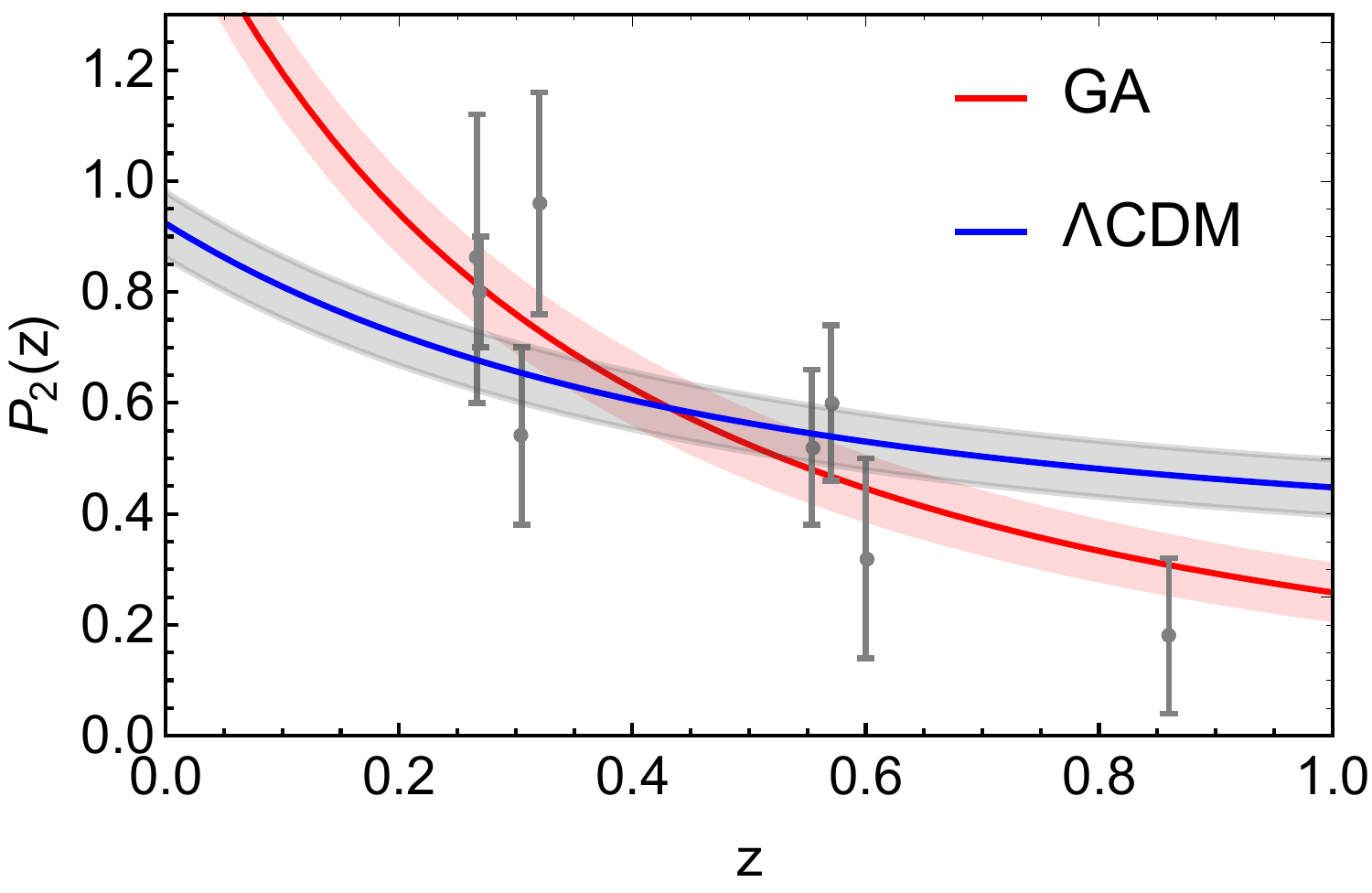}
\caption{Left: The $f\sigma_8$ data compilation along with the \lcdm best-fit (blue line) and the GA best-fit (red line). The GA reconstruction follows both the data and the \lcdm model closely. Right: The $P_2$ parameter of Ref.~\cite{Pinho:2018unz} along with the $E_\textrm{g}$ data given in Table~\ref{tab:Eg}. 
In both cases the red and grey shaded regions correspond to the $1\sigma$ confidence region for the GA and \lcdm respectively, and the actual data are shown as grey points in the background. \label{fig:fs8-P2}}
\end{figure*}

In particular, the plots of the reconstructed quantities are shown in Fig.~\ref{fig:Hz_snia} for the SnIa and $H(z)$ data, while in Fig.~\ref{fig:fs8-P2} for the $f\sigma_8$ and $E_\textrm{g}$ data. In both cases, we show both the \lcdm best-fit (blue line) and the GA best-fit (red line), while the red and grey shaded region correspond to the $1\sigma$ confidence region for the GA and \lcdm respectively and the actual data are shown as grey points in the background. The best-fit functions are then as follows:
\ba
H(z)/H_0&=&\left(1+z\left(-0.676-0.221z+0.018z^3\right)^2\right),~~~~~\\
d_\textrm{L}(z)&=&\frac{c}{H_{0}}z\left(1+z\left(0.872-0.133z-0.002z^2\right)^2\right),~~~~~\\
\frac{H_{0}}{c}d_\textrm{A}(z)&=&D_\textrm{A}(z)=\frac{z}{(1+z)^2}\left(1+z\left(0.885-0.175z+0.025z^2-0.00003z^5\right)^{2}\right),~~~~~\\
f\sigma_8(a)&=&f_0\left(a-a^4\left(-1.675+0.870a+0.001a^{2}\right)^2\right),~~~~~\\
P_2(a)&=&\left(-0.434a-0.414ae^{0.666a}-0.011a^{2.060}\right)^2,~~~~~
\ea
where $f_{0}=1.06477$, while we also find the following derived parameters
\ba
\hspace{-0.5cm}\sigma_8 &=&0.805\pm 0.246, \\
\hspace{-0.5cm}\Omega_\textrm{m,0}&=&0.254\pm 0.025,\\
\hspace{-0.5cm}\gamma_0&=&0.5549\pm 0.0003,\\
\hspace{-0.5cm}r_\textrm{sh}&=&101.873\pm 2.078~\textrm{Mpc/h},\\
\hspace{-0.5cm}w_\textrm{GA,0} &=&-0.932\pm 0.177.
\ea


Note that $E(z)=H(z)/H_0$ is reconstructed directly from the $H(z)$ data, without assuming a value of $H_0$. However, using Eq.~\eqref{eq:H0bf} we can derive a value for $H_0$ as well, which we present below, while from the $H(z)$ data we also get
\ba
\hspace{-0.5cm}q_\textrm{GA,0} &=&-0.543\pm 0.118, \\
\hspace{-0.5cm}z_\textrm{GA,tr} &=&0.641\pm 0.023,
\ea
where $z_{tr}$ is the value of the transition redshift, i.e. the redshift at which the deceleration parameter changes sign.

\begin{figure*}[!t]
\centering
\includegraphics[width = 0.49\textwidth]{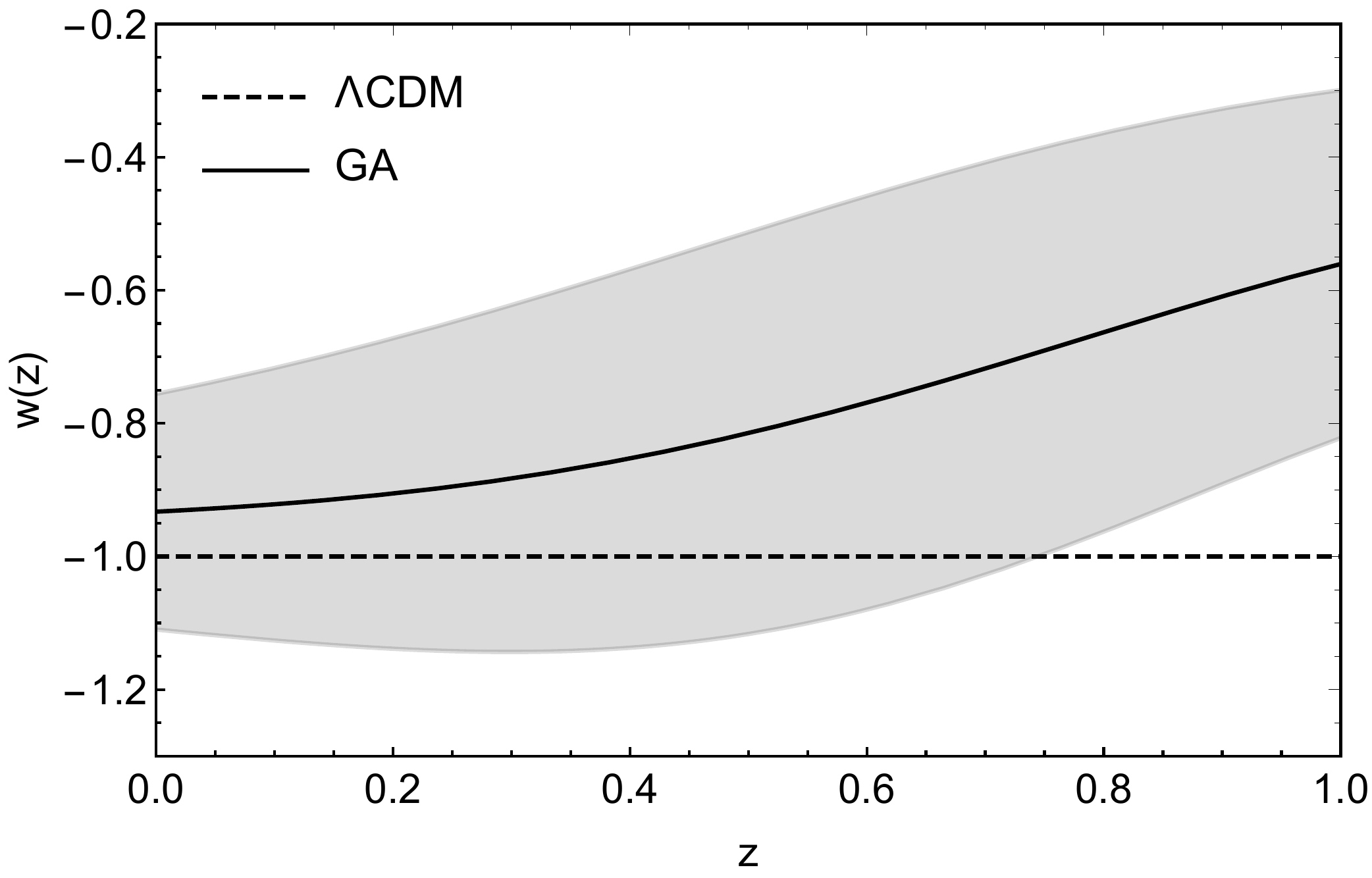}
\includegraphics[width = 0.49\textwidth]{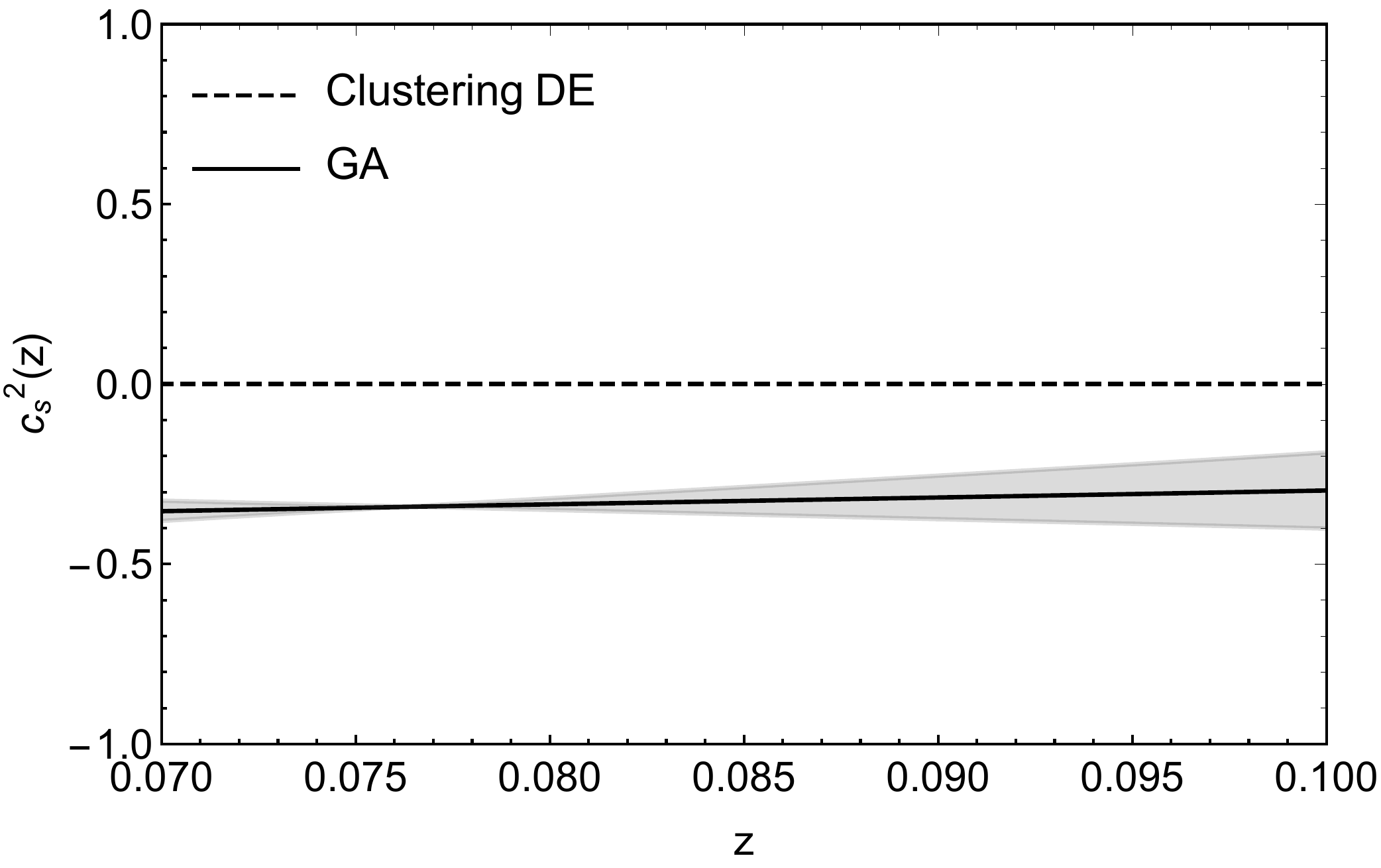}
\caption{Left: The DE equation of state $w(z)$ given by Eq.~\eqref{eq:wz}, using the GA reconstruction of the Hubble data and the value of $\Omega_\textrm{m,0}$ found from the growth data. We find that at high redshifts $(z\gtrsim 0.8)$ there is a mild $\sim1.5\sigma$ deviation from the \lcdm model. The dashed line corresponds to the theoretical prediction of the \lcdm model, while the solid black line and the grey region to the GA best-fit and the $1\sigma$ errors. Right: The adiabatic DE sound speed $c_\textrm{s,DE}^2$ given by Eq.~\eqref{eq:cs2de}. The dashed line corresponds to clustering DE with $c_s^2=0$, while the solid black line and the grey region to the GA best-fit and the $1\sigma$ errors.} \label{fig:w_F_rsh}
\end{figure*}

In Fig.~\ref{fig:w_F_rsh} we show the DE equation of state $w(z)$ given by Eq.~\eqref{eq:wz} (left panel) and the adiabatic sound speed $c_\textrm{s,DE}^2$ (right panel), where the latter is given by Eq.~\eqref{eq:cs2de}. As can be seen, the equation of state $w(z)$ is consistent with \lcdm at low redshifts, but shows a mild $2\sigma$ tension at $z\sim1$, thus hinting that deviations from the \lcdm could happen at higher redshifts as claimed in Ref.~\cite{Risaliti:2018reu}.
In the case of the adiabatic sound speed we focus on small redshifts as the earliest we can reconstruct it from the Hubble data is at $z>0.07$. As we can see, the adiabatic sound speed is evolving and is negative at the $\sim2.5\sigma$ level at $z=0.1$, which implies that DE either has a dominant non-adiabatic component at small redshifts or it should have anisotropic stress, as otherwise the matter density perturbations would be unstable \cite{Cardona:2014iba}.

\begin{figure*}[!t]
\centering
\includegraphics[width = 0.49\textwidth]{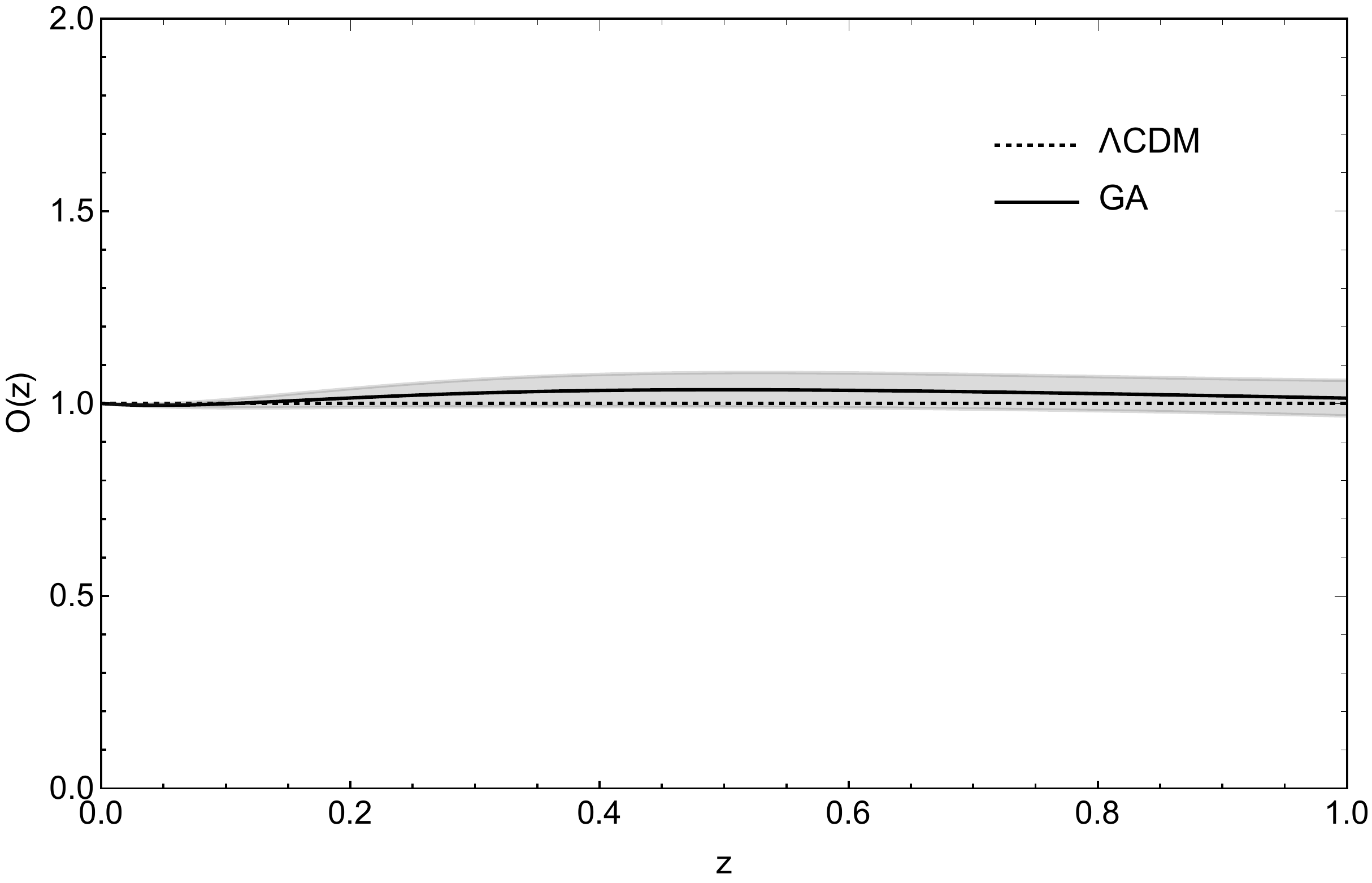}
\includegraphics[width = 0.49\textwidth]{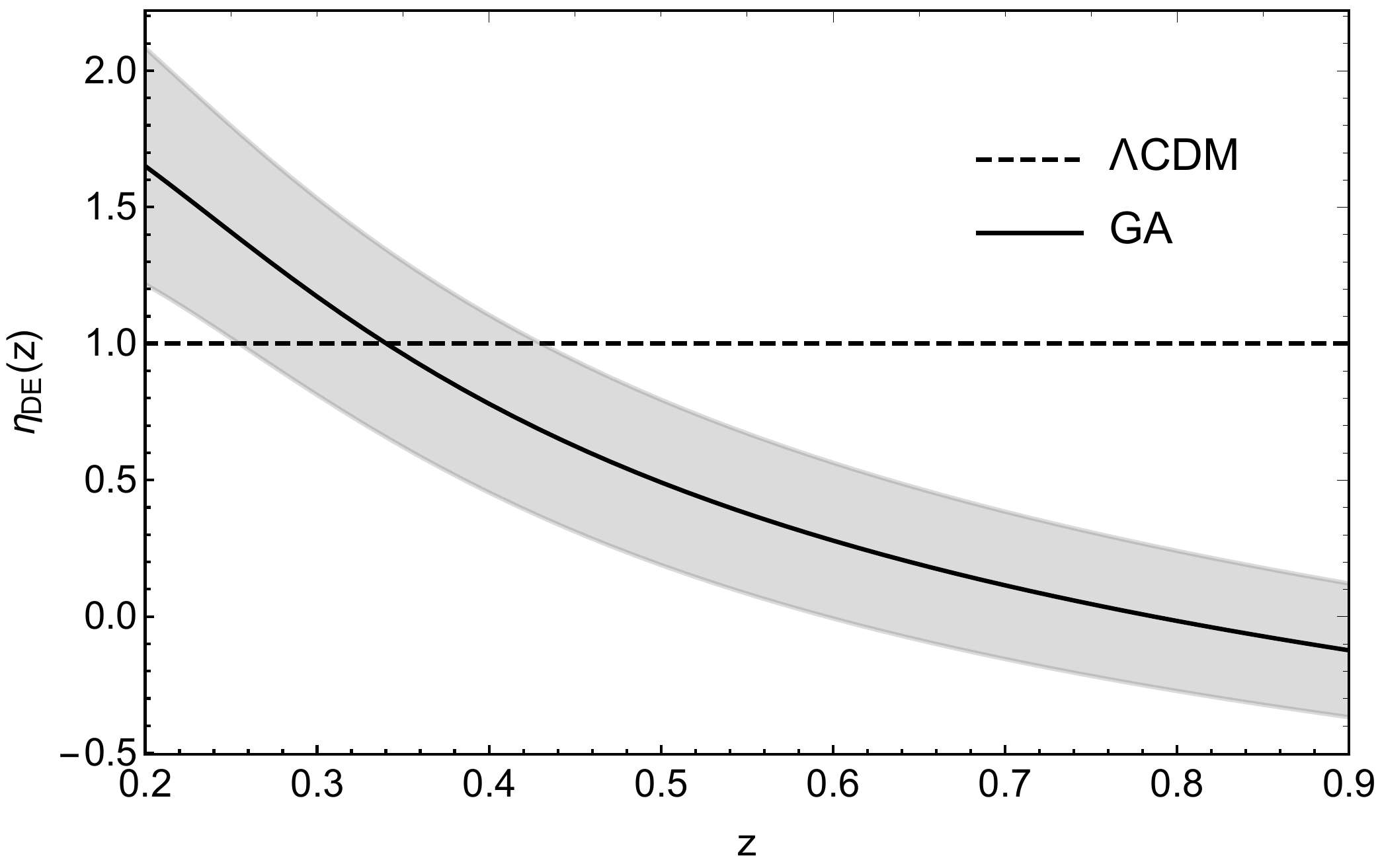}
\caption{Left: The $\mathcal{O}(z)$ test of Refs.~\cite{Nesseris:2014mfa,Nesseris:2014qca} given by Eq.~\eqref{eq:nullf}. The dashed line corresponds to the theoretical prediction of the \lcdm model (no DE anisotropic stress), while the solid black line and the grey region to the GA best-fit and the $1\sigma$ errors. We find that the test is consistent with \lcdm within the errors. Right: The anisotropic stress parameter $\eta_\textrm{DE}(z)$ given by Eq.~\eqref{eq:etade}. The dashed line corresponds to the theoretical prediction of the \lcdm model (no DE anisotropic stress), while the solid black line and the grey region to the GA best-fit and the $1\sigma$ errors. We find that there are deviations present at both low and high redshifts at the $\sim2\sigma$ and $\sim4\sigma$ level respectively.}\label{fig:O_e_rsh}
\end{figure*}

In the left panel of Fig.~\ref{fig:O_e_rsh} we show the $\mathcal{O}(z)$ test of Refs.~\cite{Nesseris:2014mfa,Nesseris:2014qca} given by Eq.~\eqref{eq:nullf}. The dashed line corresponds to the theoretical prediction of the \lcdm model, while the solid black line and the grey region to the GA best-fit and the $1\sigma$ errors. We find that the test is consistent with \lcdm within the errors. In the right panel of Fig.~\ref{fig:O_e_rsh} we show the anisotropic stress parameter $\eta_\textrm{DE}(z)$ given by Eq.~\eqref{eq:etade}. The dashed line corresponds to the theoretical prediction of the \lcdm model (no DE anisotropic stress), while the solid black line and the grey region to the GA best-fit and the $1\sigma$ errors. We find that there are deviations present at both low and high redshifts at the $\sim2\sigma$ and $\sim4\sigma$ level respectively.

\begin{figure*}[!t]
\centering
\includegraphics[width = 0.49\textwidth]{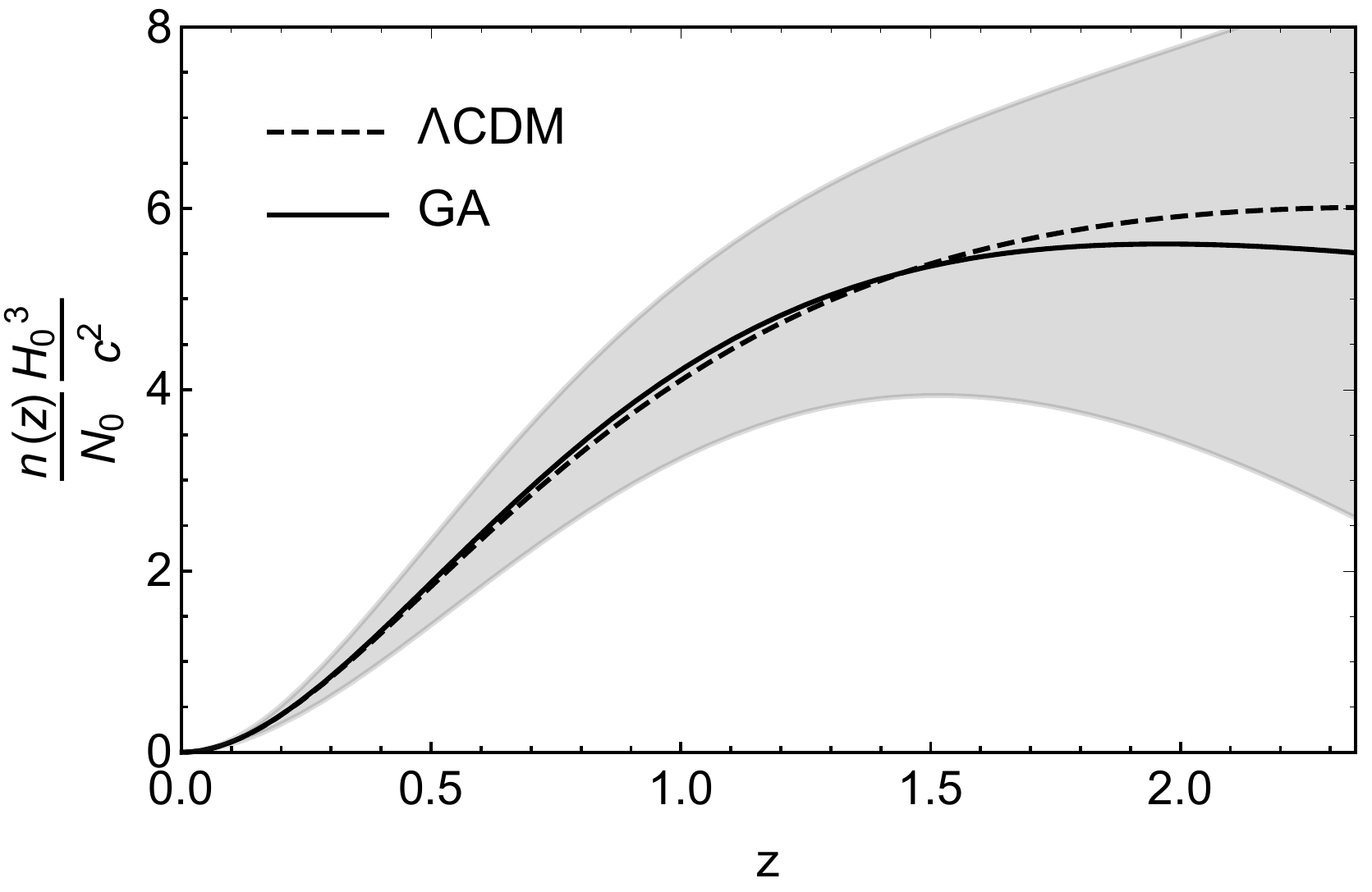}
\includegraphics[width = 0.49\textwidth]{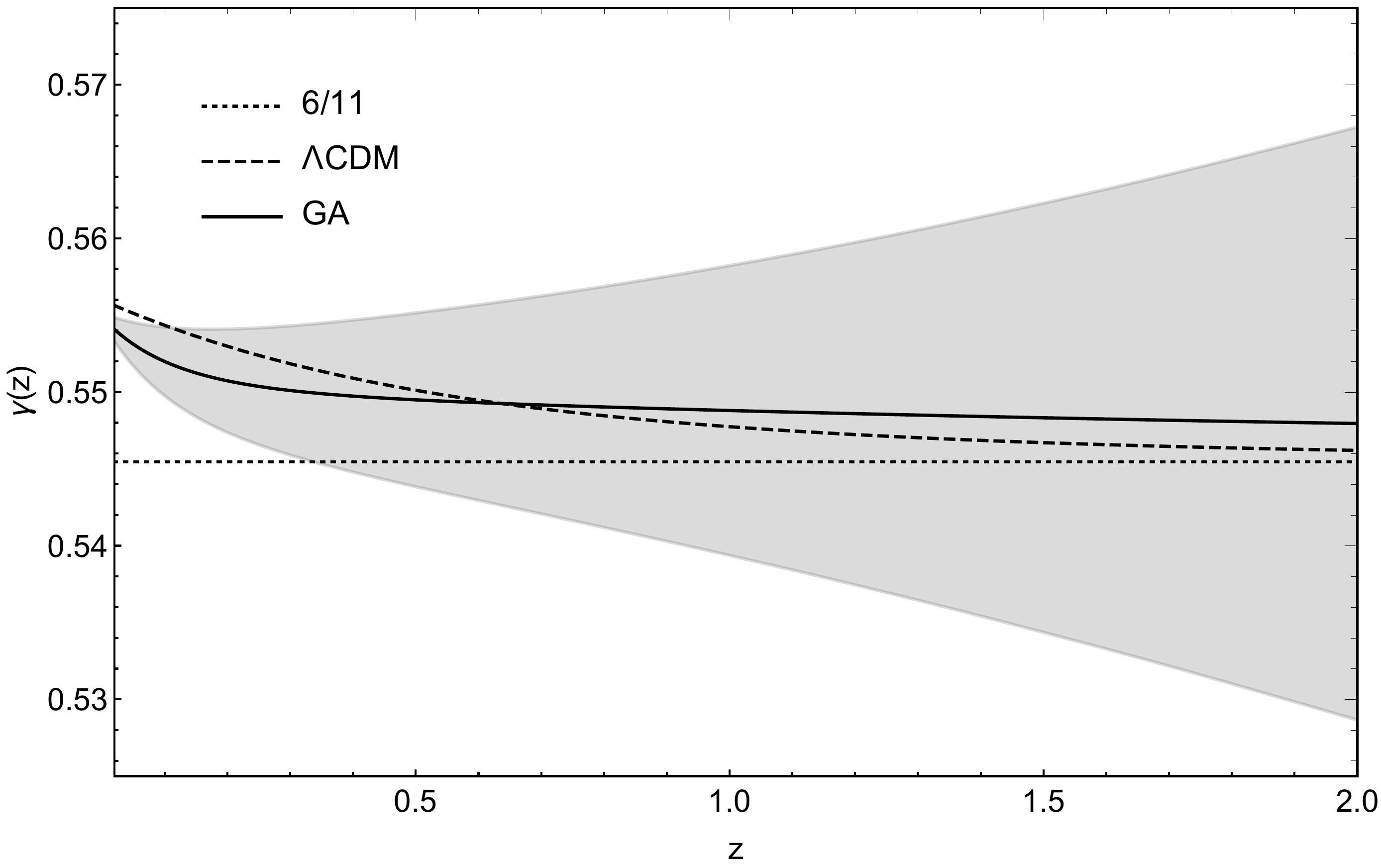}
\caption{Left: The number counts of luminous sources given by Eq.~\eqref{eq:numcounts}. The dashed line corresponds to the theoretical prediction of the \lcdm model, while the solid black line and the grey region to the GA best-fit and the $1\sigma$ errors. Right: The growth index $\gamma(z)$ of the matter density perturbations given by Eq.~\eqref{eq:gamma2}. The dashed line corresponds to the \lcdm model, the dotted line to the rough estimate $\gamma\sim \frac{6}{11}$, while the solid black line and the grey region to the GA best-fit and the $1\sigma$ errors. We find that both reconstructions are consistent with the \lcdm model.\label{fig:ncounts_gamma}}
\end{figure*}

In the left panel of Fig.~\ref{fig:ncounts_gamma} we show the number counts of luminous sources given by Eq.~\eqref{eq:numcounts}. The dashed line corresponds to the theoretical prediction of the \lcdm model, while the solid black line and the grey region to the GA best-fit and the $1\sigma$ errors. We find that the reconstructions agree with the theoretical prediction of the \lcdm model within the errors. In the right panel of Fig.~\ref{fig:ncounts_gamma} we show the growth index of the matter density perturbations $\gamma(z)$, which is given by Eq.~\eqref{eq:gamma2} and which we find to be consistent within the errors with the theoretical predictions of the \lcdm model.

The deviations found in the DE anisotropic stress reconstructed from the $E_\textrm{g}$ data and the DE equation of state $w(z)$ using the $H(z)$ data may hint either to unaccounted for systematics, possibly non-negligible radiative processes or new physics. For example, a potential source of the deviations observed with the $E_\textrm{g}$ data may be due to the lensing magnification. In Refs.~\cite{Dizgah:2016bgm,Ghosh:2018ijm} it was shown that lensing  magnification modifies both the galaxy-galaxy lensing correlations and the galaxy-galaxy correlations. As a result, lensing magnification both introduces systematic errors in the determination of $E_\textrm{g}$ and makes it bias dependent. For a more in-depth discuss of the systematics see also Ref.~\cite{Skara:2019usd}.

We should note that traditionally one would compare the $\chi^2$ per degree of freedom (dof), where the latter is traditionally defined as the number of points (36 $H(z)$ + 1048 SnIa + (4 + 3 + 2 + 2 + 1) BAO + 22 growth + 8 $E_\textrm{g}$ = 1126 points in our analysis) minus the number of free parameters of the model in question. As the GA have no free parameters, we cannot compare the dof between the GA and the \lcdm model.

\section{Conclusions \label{sec:conclusions}}
In summary, we use the Genetic Algorithms, a specific machine learning method, to reconstruct the evolution of the background history of the Universe and the matter density perturbations, based on a plethora of cosmological data including SnIa, BAO, $H(z)$, $E_\textrm{g}$ and growth rate data. The GAs can provide another method to probe underlying physical models, whilst also giving error estimates in agreement with approaches such as Fisher and MCMC \cite{Nesseris:2012tt}. In particular this has been shown to be the case with cosmological data based on several different
fiducial models such as the \lcdm model and modified gravity theories \cite{Hogg:2020ktc}, but also models that violate the distance duality relation \cite{Martinelli:2020hud}.

Using then the GA and the cosmological data described in Sec.~\ref{sec:data}, we find that there is a $\sim2\sigma$ deviation of $w(z)$ from -1 at high redshifts, the adiabatic sound speed $c_\textrm{s,DE}^2$ is evolving and is negative at the $\sim2.5\sigma$ level at $z=0.1$, while using the $E_\textrm{g}$ data we find a $\sim2\sigma$ deviation of the anisotropic stress $\eta_\textrm{DE}(z)$ from unity at low redshifts and $\sim4 \sigma$ at high redshifts, thus suggesting the presence of significant deviations from the \lcdm model. The reconstructions of these quantities in terms of the redshift $z$, along with the $1\sigma$ error regions, is shown in Figs.~\ref{fig:w_F_rsh} and \ref{fig:O_e_rsh}.

As mentioned earlier, our approach has been validated in previous works via the use of mocks, see Refs.~\cite{Hogg:2020ktc,Martinelli:2020hud}, so clearly then the aforementioned deviations from the \lcdm model present a problem as they hint towards two possibilities, either the presence of unaccounted for systematics and possibly non-negligible radiative processes, as might be the case for the $E_\textrm{g}$ data, or new physics in the form of modifications of gravity. The latter case is quite plausible, as the deviations come from very different data sets with very different systematics, i.e. the equation of state $w(z)$ and $c_\textrm{s,DE}^2$ from the $H(z)$ data, the growth index from the growth data coming from the RSD measurements and the $\eta_\textrm{DE}(z)$ from the $E_\textrm{g}$ data.

Specifically, the fact that the adiabatic sound speed $c_\textrm{s,DE}^2$ is both evolving and negative, implies that the DE perturbations would be unstable unless there exists either a strong anisotropic stress, coming for example from some modification of gravity, so that the total effective sound speed is positive, as shown in Ref.~\cite{Cardona:2014iba} or a non-adiabatic DE component \cite{Arjona:2020yum}. In particular, using the effective fluid approach it can be shown that in $f(R)$ models, like the Hu-Sawicki or the designer model, the sound speed of the effective DE fluid is negative and the matter perturbations are stable due to the anisotropic stress \cite{Arjona:2018jhh}, hence lending more support to modified gravity scenarios.

A possible caveat in our analysis is that there is some overlap between the BAO and $H(z)$ data, which could induce spurious correlations between the two data sets. Since unfortunately, we do not have access to the covariance matrices, we are not able to take this covariance into account, thus possibly underestimating the errors in our analysis, hence we analyze them separately. This is also the case for the $E_\textrm{g}$ and the growth rate data, which come from overlapping surveys, however again we do not have the covariance matrices.

However, our approach is completely agnostic as we made no assumptions about the nature of DE or the spatial curvature of the Universe during the fitting of our data. This is one of the main advantages of our ML approach compared to other traditional or non-parametric methods such as cosmography which suffers from convergence issues at high redshifts or Gaussian processes that assume a fiducial model. As the GA can provide model-independent reconstructions of key parameters that describe DE, then if indeed there are no systematics in the data, the observed model-independent deviations from \lcdm could point to the existence of new physics. The possibility of such an exciting prospect could be further strengthened by the upcoming cosmological surveys like LSST \cite{Abell:2009aa}.


\begin{acknowledgments}
The authors thank G.~Ballesteros and L.~Perivolaropoulos for useful discussions and acknowledge support from the research project PGC2018-094773-B-C32 and the Centro de Excelencia Severo Ochoa Program SEV-2016-0597. S.~N. also acknowledges support from the Ram\'{o}n y Cajal program through Grant No. RYC-2014-15843.
\end{acknowledgments}

\section*{Numerical codes}
The GA codes used in our paper are freely available at \href{https://github.com/snesseris}{https://github.com/snesseris} and \href{https://github.com/RubenArjona}{https://github.com/RubenArjona}.

\bibliographystyle{JHEP}
\bibliography{ML_LSS}

\end{document}